\documentclass[12pt]{iopart}

\expandafter\let\csname equation*\endcsname\relax
\expandafter\let\csname endequation*\endcsname\relax
\usepackage{amsmath}
\usepackage{xcolor}
\usepackage{graphicx} % Required for inserting images
\usepackage{url}
\usepackage[T1]{fontenc}

\newcommand{\gp}{{$g^{\prime}$}}
\newcommand{\rp}{{$r^{\prime}$}}
\newcommand{\ip}{{$i^{\prime}$}}

\begin{document}

\title[ROME/REA Survey DR1]{ROME/REA: Three-year, Tri-color Timeseries Photometry of the Galactic Bulge}

\author[0000-0001-6279-0552]{R.A.~Street}
\address{Las Cumbres Observatory, 6740 Cortona Drive, Suite 102, Goleta, CA 93117, USA}
\ead{rstreet@lco.global}

\author[0000-0002-6578-5078]{E.~Bachelet}
\address{IPAC, Mail Code 100-22, Caltech, 1200 E. California Blvd., Pasadena, CA 91125, USA}

\author[0000-0001-8411-351X]{Y.~Tsapras}
\address{Astronomisches Rechen-Institut, M{\"o}nchhofstr. 12-14, D-69120 Heidelberg,
Germany}

\author{M.P.G.~Hundertmark}
\address{Astronomisches Rechen-Institut, M{\"o}nchhofstr. 12-14, D-69120 Heidelberg,
Germany}

\author{V.~Bozza}
\address{Dipartimento di Fisica "E.R. Canianiello", Universit{\`a} di Salerno, Via Giovanni Paolo II 132, 84084, Fisciano, Italy}
\address{Istituto Nazionale di Fisica Nucleare, Sezione di Napoli, Via Cintia, 80126 Napoli, Italy}

\author[0000-0002-1583-6519]{D.M.~Bramich}
\address{Center for Astrophysics and Space Science, New York University Abu Dhabi, PO Box 129188,
Saadiyat Island, Abu Dhabi, United Arab Emirates}

\author{A.~Cassan}
\address{Institut d’Astrophysique de Paris, Sorbonne Universit\'e, CNRS, UMR 7095, 98 bis bd Arago, 75014 Paris, France}

\author[0000-0002-3202-0343]{M.~Dominik}
\address{University of St Andrews, Centre for Exoplanet Science, SUPA School of Physics \&
Astronomy, North Haugh, St Andrews, KY16 9SS, United Kingdom}

\author{R. Figuera Jaimes} 
\address{Millennium Institute of Astrophysics MAS, Nuncio Monsenor Sotero Sanz 100, Of. 104, Providencia, Santiago, Chile}
\address{Instituto de Astrof\'isica, Facultad de F\'isica, Pontificia Universidad Cat\'olica de Chile, Av. Vicu\~na Mackenna 4860, 7820436 Macul, Santiago, Chile}

\author{K. Horne}
\address{Centre for Exoplanet Science, SUPA, School of Physics \& Astronomy, University of St Andrews, North Haugh, St Andrews KY16 9SS, UK}

\author[0000-0001-8317-2788]{S.~Mao}
\address{Department of Astronomy, Tsinghua University, Beijing, China 100084}

\author{A.~Saha}
\address{National Optical Astronomy Observatory, 950 North Cherry Ave., Tucson, AZ 85719, USA}

\author{J.~Wambsganss}
\address{Zentrum f{\"u}r Astronomie der Universit{\"a}t Heidelberg, Astronomisches Rechen-Institut, M{\"o}nchhofstr. 12-14, 69120 Heidelberg, Germany}
\address{International Space Science Institute (ISSI), Hallerstra{\ss}e 6, 3012 Bern, Switzerland}

\author[0000-0001-6000-3463]{Weicheng~Zang}
\address{Department of Astronomy, Tsinghua University, Beijing, China 100084}

\vspace{10pt}
\begin{indented}
\item[]December 2023
\end{indented}

\begin{abstract}
The ROME/REA (Robotic Observations of Microlensing Events/Reactive Event Assessment) Survey was a Key Project at  Las Cumbres Observatory (hereafter LCO) which continuously monitored 20 selected fields (3.76 sq.deg.) in the Galactic Bulge throughout their seasonal visibility window over a three-year period, between March 2017 and March 2020.  Observations were made in three optical passbands (SDSS$-g^{\prime}$, $-r^{\prime}$, $-i^{\prime}$), and LCO's multi-site telescope network enabled the survey to achieve a typical cadence of $\sim$10\,hrs in $i^{\prime}$ and $\sim$15\,hrs in $g^{\prime}$ and $r^{\prime}$.   In addition, intervals of higher cadence ($<$1\,hr) data were obtained during monitoring of key microlensing events within the fields.  This paper describes the Difference Image Analysis data reduction pipeline developed to process these data, and the process for combining the photometry from LCO's three observing sites in the Southern Hemisphere. The full timeseries photometry for all $\sim$8\, million stars, down to a limiting magnitude of $i\sim18$\,mag is provided in the data release accompanying this paper, and samples of the data are presented for exemplar microlensing events, illustrating how the tri-band data are used to derive constraints on the microlensing source star parameters, a necessary step in determining the physical properties of the lensing object.  The timeseries data also enables a wealth of additional science, for example in characterizing long-timescale stellar variability, and a few examples of the data for known variables are presented.
\end{abstract}

%\maketitle

\section{Introduction}
In center of the Galactic Bulge there is a window (centered at RA$\sim$18, 	Dec$\sim$-28.5, and $\sim$few tens of degrees wide) through which the observer can explore several stellar populations in the Milky Way Disk, Bulge and Halo.  This region has been the subject of consistent photometric monitoring for $\sim$30\,yrs, due to the high rate of microlensing events that occur within this region \cite{Mroz2019}.  

Microlensing occurs when a foreground massive body, called the lens, crosses the observer's line of sight to a background luminous source.  The gravity of the lens deflects the source star's light, causing the observer to see a gradual brightening and fading of the source as the objects move through alignment.  Since these objects are normally unrelated, these events are true transients, and inherently rare (optical depth $<4\times10^{-6}$ \cite{Mroz2019}), so surveys seeking to discover them have traditionally concentrated on crowded regions in order to monitor as many stars as possible \cite{Tsapras2018}.  This means that relatively high spatial resolution ($\sim$1\,arcsec/pixel or better) imaging is optimal to resolve the overlapping stellar Point Spread Functions (PSFs).  

Although challenging to discover, microlensing events are scientifically valuable as they provide the means to measure the masses of objects that would otherwise be too faint to observe, including free-floating planets \cite{Sumi2023, Koshimoto2023, Gould2022}, and even isolated compact objects such as black holes and neutron stars \cite{Lam2023, Sahu2022}.  Planetary, Brown Dwarf and stellar companions of lensing stars can betray their presence by causing short-lived ($\sim$hours--days) `anomalous' deviations to the otherwise-smooth lensing light curve \cite{Mao1991}.  Microlensing events are most sensitive to planets between $\sim$1--10\,AU from their host stars, thought to be a key region in planet formation around the so-called snowline where ices condense in circumstellar disks and planetesimal formation is favored in a wide range of circumstances \cite{Drkazkowska2017}.  This area of parameter space is practically difficult or time-consuming to explore with other planet-hunting methods.  To date there have been 5539 confirmed discoveries of planetary systems\footnote{NASA Exoplanet Archive https://exoplanetarchive.ipac.caltech.edu/}, of which 204 were detected from their microlensing signatures.  The majority of these events were detected by microlensing surveys such as the Optical Gravitational Lensing Experiment (OGLE) \cite{Udalski1992ogleproject}, Microlensing Observations in Astrophysics (MOA) \cite{Sako2008moaproject} and the Korean Microlensing Telescope Network (KMTNet) \cite{Park2012KMTNetproject}, with follow-up observations contributed for high priority events by ground-based follow-up teams such as MicroFUN \cite{Yoo2004}, PLANET \cite{PLANET2002}, RoboNet \cite{RoboNet2009} and MiNDSTEp \cite{MiNDSTEp}.  Space-based facilities, notably the {\em Spitzer} Space Telescope \cite{Dong2007, Udalski2015} and the {\em K2} Campaign 9 \cite{Henderson2016}, have also provided valuable photometric constraints on microlensing parallax, while more recently the {\em Gaia} Mission has delivered timeseries photometry and astrometry \cite{Gaia_DR3}.  Due to a limitation of {\em Gaia}'s pipeline, it can measure up to $\sim$1,050,000 objects per square degree.  Fields with a higher stellar density, like the Bulge, must downlink the fullframe images in order to build a more complete catalog, which is rarely done \cite{GaiaCollaboration2023}.  Instead, the {\em Gaia} catalog for Bulge fields is restricted to bright stars only.  There have also been some infrared surveys of the region including from the UK InfraRed Telescope (UKIRT) \cite{Shvartzvald2017} and the VISTA Variables in the Via Lactea surveys (VVV and VVVX \cite{VVV2010}) on the 4.1\,m Visible and Infrared Survey Telescope for Astronomy.  Many of these IR surveys targeted regions of high extinction ($b\sim0^{\circ}$) and so do not fully overlap the optical survey footprints. 

Despite the wealth of data on the region, each survey was designed with specific observational constraints, and only some of the resulting data products are publicly accessible. Most microlensing surveys prioritize high-cadence (from every $\sim$10\,min to $<$1\,d$^{-1}$) photometry in a single passband, in order to fully sample fleeting planetary anomalies.  Data in other passbands are obtained with a cadence of $\sim$1\,d or lower.  A bulk download of the single-band MOA photometry is public\footnote{https://exoplanetarchive.ipac.caltech.edu/docs/MOAMission.html}, including extended baseline photometry between 2006--2014.  Photometry from the $H-$ and $K-$band UKIRT survey, obtained between 2015–-2019, has also been released\footnote{https://exoplanetarchive.ipac.caltech.edu/cgi-bin/TblSearch/\\nph-tblSearchInit?app=ExoTbls\&config=ukirttimeseries}.  
The {\em Gaia} and {\em VVV} surveys provide photometry with multi-year baselines but typically have lower cadence than dedicated microlensing surveys, with multiple days between visits to a field as opposed to minutes--hours intervals.   Almost all of the ground-based surveys have been conducted from a single-site so their light curves have day-gaps, except for KMTNet, which operates telescopes in Chile, South Africa and Australia.  KMTNet have released selected subsets of their photometry. 

Multi-band timeseries photometry is valuable in microlensing because it can be used to infer the spectral type, and hence the angular radius of the source star in microlensing events (e.g. \cite{Bachelet2022, Rybicki2022}).  From this an independent estimate of the distance to the source can be inferred, which, when combined with the parameters of the microlensing lightcurve model, allows the mass and distance to the lens to be inferred.  Spectroscopy is also sometimes used but as most microlensing sources are $V>16$\,mag, this can be challenging.  Since only the lensed source is magnified during the event, the source star's flux can be distinguished from any blended neighbors provided observations are obtained at different magnifications.  This means that regular observations in at least two filters are required, but tri-band timeseries provides more constraints on Spectral Energy Distribution of the source, and allows the flux from the source to be distinguished from that of blended stars using a linear-regression procedure outlined in \cite{Street2019}.  As this is independent of the fitted microlensing model, it provides a valuable check on the source and blend flux parameters normally fitted as part of microlensing models.  
The ROME/REA Project \cite{Tsapras2019} was designed to deliver multi-year, multi-band optical timeseries photometry of a large set of microlensing events in the central Galactic Bulge, taking advantage of the multi-site Las Cumbres Observatory Telescope Network (LCO) to provide imaging every few hours.  This is highly complementary to data from other contemporaneous surveys.  In addition to microlensing events, the data from this survey include astrophysical variables of all kinds, from eclipsing binaries to RR~Lyrae.  In this paper, we present the a data release of the full photometric timeseries data from the ROME/REA survey.  By publishing the {\em entire} catalog, rather than selected timeseries of known variables, the data may also be used to train and test machine learning classification algorithms with real-world variety in variability and data quality, as the observing strategy mimics the multi-filter imaging expected from the Rubin Observatory's Legacy Survey of Space and Time.  

In Sections~\ref{sec:romerea} and \ref{sec:observations} we describe the survey design and the set of observations realized in practice.  Section~\ref{sec:pipeline} describes the new open-source Difference Image Analysis (DIA) pipeline developed to process these data, together with the method used to calibrate the photometry from different telescopes.  In Section~\ref{sec:results} we present exemplar results from the project and demonstrate the science that can be done with the data, while Section~\ref{sec:data_products} provides a full description of the resulting data products.  

\section{The ROME/REA Project}
\label{sec:romerea}

The ROME/REA Project was an LCO Key Project that conducted observations between 2017--2020.  It consisted of two elements: a regular survey of selected fields in the Galactic Bulge plus additional, higher-cadence observations made in response to alerts of microlensing events within those fields.   In this manner, the project took advantage of the unique features of the LCO telescope network, by using its geographically distributed sites to maintain long-baseline around-the-clock monitoring of the fields, while using the multiple telescopes at each site to simultaneously coordinate targeted observations of high priority events.  Tsapras et al.(2019) \cite{Tsapras2019} provides a full description of the project as planned, so here we discuss how the project was realized in practice.  

\subsection{Instrumentation}
The LCO network currently consists of 2\,m, 1\,m and 0.4\,m telescopes located at 7 observatory sites around the world \cite{LCONetwork}.  The entire network is robotically operated, with observations automatically scheduled through LCO's dynamic scheduling software \cite{Lampoudi2015}. Telescopes in each aperture class are as identical as possible in design and instrumentation; for example all of the 1\,m telescopes in the network support 4k$\times$4k Sinistro imaging cameras, which offer a 26$\times$25\,arcmin field of view and the same complement of filters, including Johnson-Cousins, Bessell, Sloan Digital Sky Survey and PanSTARRS-Y filtersets.  Full information about the Sinistro cameras can be found in LCO's website\footnote{https://lco.global/observatory/instruments/sinistro/}.  The consistency of instrumentation at each site, combined with several sites hosting multiple telescopes in the same (and sometimes different) aperture class, plus LCO's rapid scheduling enables a range of unique observing strategies.  A single object can be monitored around the clock for the whole of its seasonal visibility period, with observations being automatically transferred to different telescopes or sites to mitigate for poor weather or technical downtime at an individual facility.  For transient events like microlensing, LCO's ability to respond within $\sim$10\,mins to observation requests permits rapid response observations of high priority events with one telescope while simultaneously maintaining regular survey-mode observations on a different telescope at the same site.  

The Galactic Bulge is primarily visible to LCO's Southern Ring of 1\,m telescopes at Cerro Tololo in Chile, Sutherland in South Africa and Siding Spring in Australia.  All three sites host three 1\,m telescopes, each with a Sinistro camera.  To maximize consistency of data acquisition, the ROME survey-mode observations were always scheduled on the Dome-A telescopes at each site.  More flexibility was granted for the REA-mode observations made in response to alerts to maximize the speed of response, and observations were conducted on Domes B and C at each site.  

While we attempted to consistently use the same set of instruments, inevitably routine network maintenance resulted in some changes.  For example, the fa03 camera was moved from Dome B in Chile to Dome C in July 2017, as issues with a different camera required it to be replaced, and swapping the cameras over was the most efficient way to bring both telescopes back on sky as soon as possible.  For data reduction purposes, the data from a single combination of field pointing, site, telescope, camera and passband is treated as a separate dataset.  The pipeline can also be configured to recognize alternative binnings of image data to distinguish datasets, but as the data for ROME/REA were taken with a single binning (the LCO default of 1$\times$1), this option was not applied for this work. 
 The full list of instruments used for the project are presented in Table~\ref{tab:telescopes}.  

\begin{table}[]
    \centering
    \begin{tabular}{|l|l|l|l|}
    \hline
      Site (site code)              &  Dome  &  Telescope  & Instrument\\
      \hline
      Cerro Tololo, Chile (LSC)     &  A     & 1m0-05       & fa15\\
      Cerro Tololo, Chile (LSC)     &  B     & 1m0-09       & fa03\\
      Cerro Tololo, Chile (LSC)     &  C     & 1m0-04       & fa03$*$\\
      Sutherland, South Africa (CPT)&  A     & 1m0-10       & fa16\\
      Sutherland, South Africa (CPT)&  B     & 1m0-13       & fa14\\
      Sutherland, South Africa (CPT)&  C     & 1m0-12       & fa06\\
      Siding Spring, Australia (COJ)&  A     & 1m0-11       & fa12\\
      Siding Spring, Australia (COJ)&  B     & 1m0-03       & fa11\\
      \hline
    \end{tabular}
    \caption{Summary of the telescopes and instruments used to make observations for the ROME/REA Project\\
    * The fa03 instrument was transferred from Dome B to Dome C during the ROME/REA project.}
    \label{tab:telescopes}
\end{table}

\subsection{Field Selection}
\begin{figure}
\begin{centering}
\begin{tabular}{c|c}
    \includegraphics[width=0.5\textwidth]{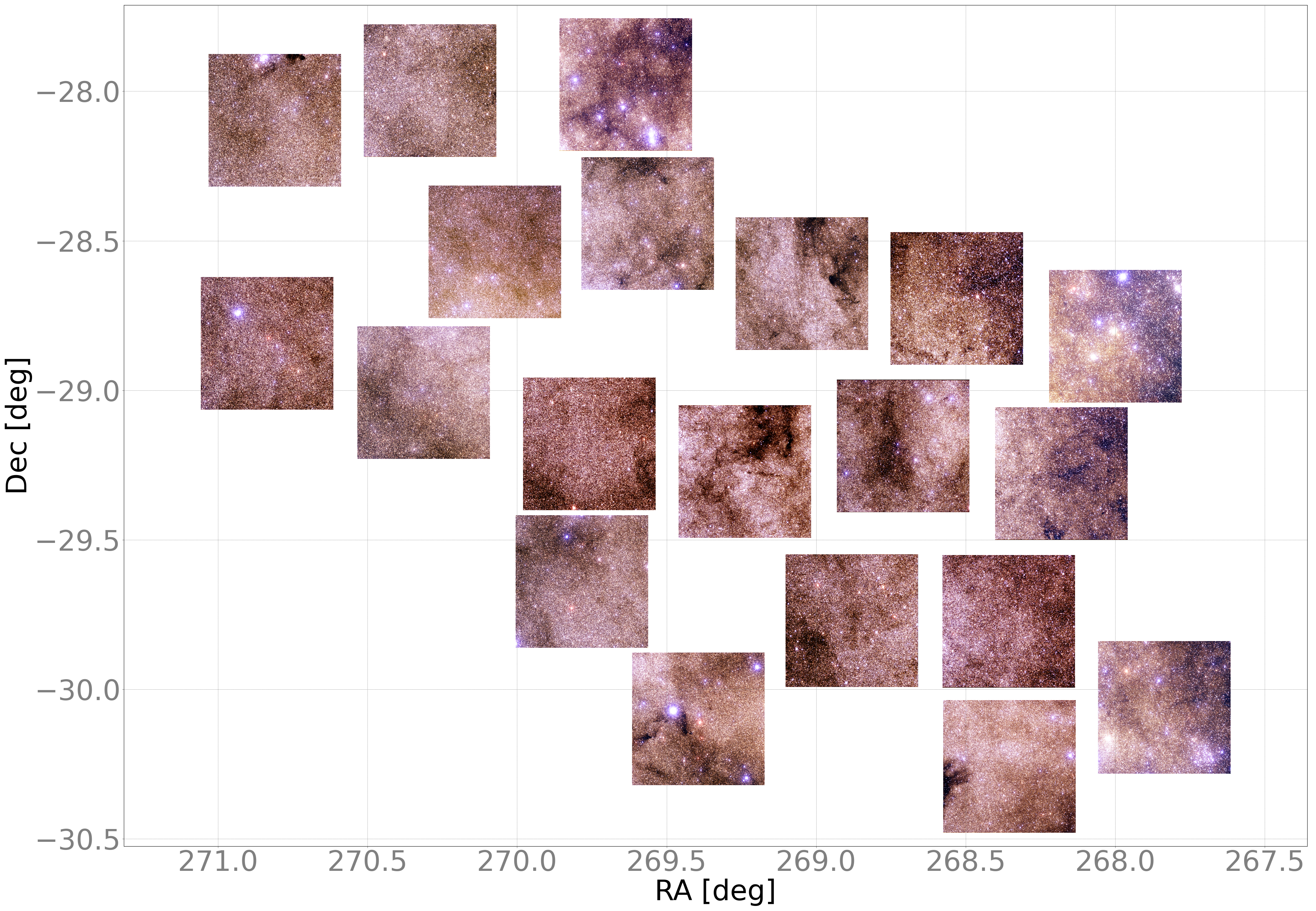} &
    \includegraphics[width=0.5\textwidth]{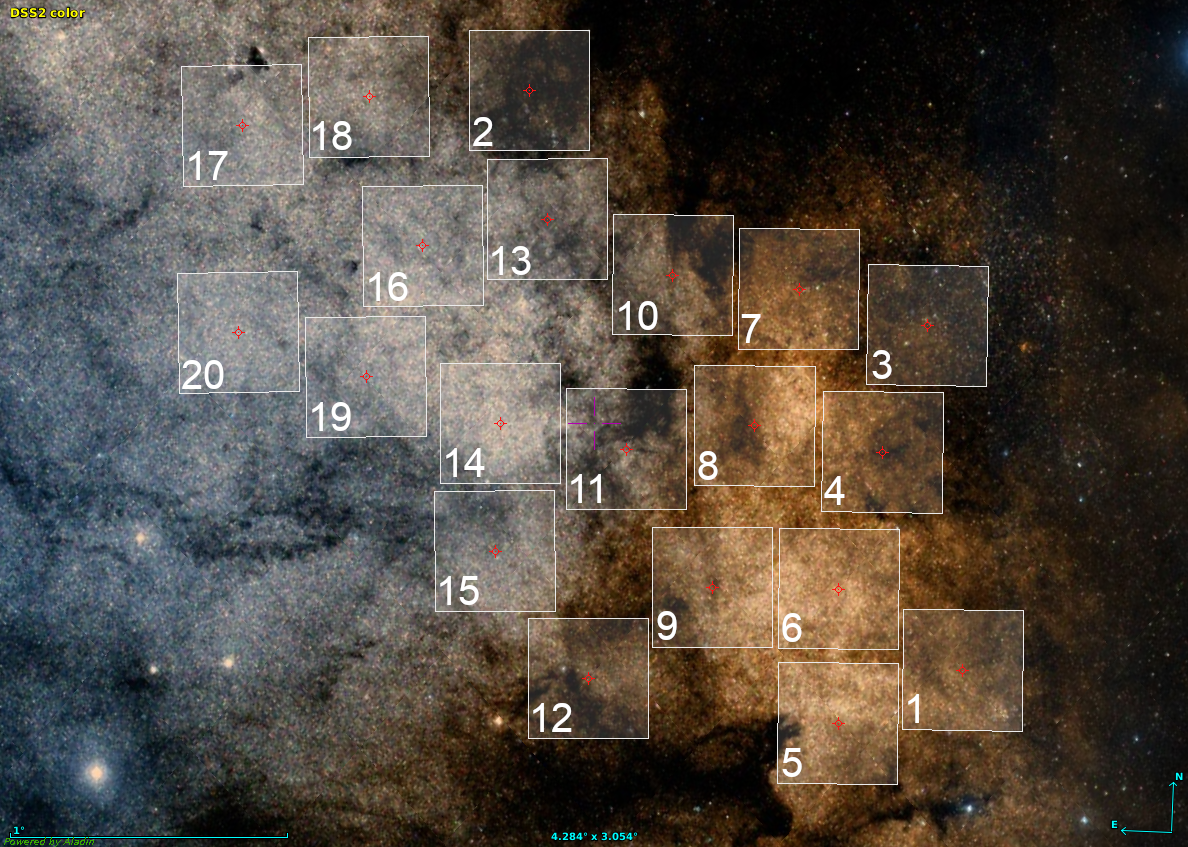} \\
    \includegraphics[width=0.4\textwidth]{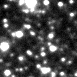} & \\
\end{tabular}
\caption{The spatial distribution of the ROME survey fields.  (Top left) Mosaic of ROME reference images, combining data in $g^{\prime}$, $r^{\prime}$ and $i^{\prime}$ to highlight the variable extinction in the fields. (Top right) Field layout overlaid on a 4.285$\times$3.054$^{\circ}$ DSS2 color image showing the fields in context of the wider galactic structure. (Bottom) Zoom into the central 30\,arcsec of the SDSS-i$^{\prime}$ reference image for ROME-FIELD-01, showing the resolved pixels across the stellar Point-Spread-Function. 
 \label{fig:survey_map}}
\end{centering}
\end{figure}

The survey fields for ROME/REA were selected (as described in \cite{Tsapras2019}), from the region in the central Galactic Bulge where the microlensing rate is highest.  A total of 20 fields were observed, covering a total area of 3.76\, sq. deg., based on a trade off between covering as large an area as possible while still ensuring each field is observed multiple times per night from sites around the network.  The pointings chosen were not contiguous, since extinction can vary by more than 1\,mag as a function of position in the Galactic Bulge.  Taking the field of view of the Sinistro cameras into account, the field pointings were adjusted to maximize the total number of stars in the survey.  They were also adjusted to avoid very bright stars ($V<7$\,mag) wherever possible, to minimize the fraction of the detectors that would be affected by column bleeds. The resulting fields are centered at approximately RA=17:57:20.7, Dec=-29:07:05.0, and lie within a radius of $\sim$2.06$^{\circ}$ of that location. A summary of the fields, and the data acquired, is presented in Table~\ref{tab:fields}.  Figure~\ref{fig:survey_map} illustrates the spatial locations of the fields.  Importantly, the same field pointings were used for REA as well as ROME-mode observations.  REA observations were conducted for  events brighter than $V\sim17$\,mag that were alerted and identified to lie within the ROME fields.  By pointing at the survey field, rather than directly at a specific event, REA contributed additional observations to the light curves of all other stars in the survey, as well as to the targeted event.  

\begin{table}[]
    \centering
    \begin{tabular}{|l|l|l|c|c|c|c|}
        \hline
       Field name  &  RA            & Dec           & N stars & \multicolumn{3}{c|}{N observations} \\
                   & \multicolumn{2}{c|}{J2000.0}   &         & \gp        & \rp           & \ip\\
      \hline
      ROME-FIELD-01 & 17:51:20.61 & -30:03:38.94    & 403366  & 516 & 616 & 1082 \\
      ROME-FIELD-02 & 17:58:32.82 &-27:58:41.76     & 353000  &  400    &  480    &  739 \\
      ROME-FIELD-03 & 17:52:00.01 & -28:49:10.41    & 350998  &  403  &   497  & 972 \\
      ROME-FIELD-04 & 17:52:43.24 & -29:16:42.65    & 396539  &  422  &  518 & 1034 \\
      ROME-FIELD-05 & 17:53:25.04 & -30:15:28.21    & 396897  & 420 & 531 & 1158 \\
      ROME-FIELD-06 & 17:53:25.47 & -29:46:22.73    & 401336  & 390 & 454 & 1205 \\
      ROME-FIELD-07 & 17:54:07.10 & -28:41:37.35    & 437155  & 411 & 506 & 1123 \\
      ROME-FIELD-08 & 17:54:50.34 & -29:11:12.21    & 407645  & 406 & 497 & 1483 \\
      ROME-FIELD-09 & 17:55:31.47 & -29:46:13.68    & 462737  & 392 & 497 & 1302 \\
      ROME-FIELD-10 & 17:56:11.64 & -28:38:38.64    & 418142  & 374 & 463 & 757 \\
      ROME-FIELD-11 & 17:56:57.32 & -29:16:18.01    & 430265  & 425 & 526 & 1644 \\
      ROME-FIELD-12 & 17:57:34.75 & -30:05:57.25    & 369112  & 422 & 511 & 984 \\
      ROME-FIELD-13 & 17:58:15.29 & -28:26:32.04    & 392990  & 390 & 480 & 774 \\
      ROME-FIELD-14 & 17:59:02.12 & -29:10:46.57    & 422179  & 463 & 578 & 1058 \\
      ROME-FIELD-15 & 17:59:08.06 & -29:38:21.86    & 414096  & 292 & 347 & 544 \\
      ROME-FIELD-16 & 18:00:18.00 & -28:32:15.21    & 445014  & 437 & 583 & 1764 \\
      ROME-FIELD-17 & 18:03:14.40 & -28:05:52.20    & 465204  & 435 & 528 & 990 \\
      ROME-FIELD-18 & 18:01:09.81 & -27:59:54.97    & 431414  & 417 & 509 & 870 \\
      ROME-FIELD-19 & 18:01:15.06 & -29:00:30.33    & 478540  & 415 & 496 & 1963 \\
      ROME-FIELD-20 & 18:03:20.82 & -28:50:35.37    & 498284  & 440 & 539 & 1178 \\
      \hline
      Total         &             &                 & 8374913 & 8270 & 10156 & 22623 \\
      \hline
      \hline
    \end{tabular}
    \caption{Summary of the fields surveyed for the ROME/REA Project, including the number of stars and observations that passed data quality checks.}
    \label{tab:fields}
\end{table}

\section{Observations and Data}
\label{sec:observations}
For each field, survey mode observation requests were submitted as a set of \gp,\rp,\ip\ exposures, with 2$\times$300\,s in all bands, to be repeated at a cadence of nominally 7\,hrs, but allowing a `jitter' of 7\,hrs.  This parameter allows the LCO scheduling algorithm flexibility in sequencing repeated observations, and the exposure time was determined from test observations to provide $\sim$1--few hundreths mag photometry for stars in our target range of \ip$\sim$14--17\,mag at event peak.  All survey observations were assigned a fixed `Intra-Proposal Priority' (IPP) factor of 1.05, which is the default and equates to no extra weighting in the LCO scheduler, relative to other observation requests.  The survey mode observations were submitted to the telescopes in Dome A at each of the three Southern hemisphere sites, and this observing strategy was continued for as long as the Galactic Bulge was visible, each year of the 3-yr project.  Constrained by the annual visibility of the Bulge, observations were performed in seasons spanning from March to October each year.  

Targets for reactive-mode observations were selected automatically by our TArget Prioritization algorithm (TAP) \cite{Hundertmark2018}. Although these observations were centred on the pointing of the survey fields, these observations were given exposure times tailored to the current brightness of the targeted event to avoid saturating bright targets.  This was predicted based on real-time analysis of the event lightcurve, so REA-mode observations were updated daily, with the exposure times calculated from a function based on currentmagnitude.  REA-mode observations were conducted in \ip-band, to ensure high cadence monitoring over the peak of the events and to provide coverage of any anomalous features. Active events for REA were required to have expected Einstein crossing timescales of $<300\,\rm{d}$ and a predicted magnification of $>1.34$, corresponding to the lensing zone within the Einstein radius (the characteristic angular radius around the lens where images of the source star form due to the deflection of light).  The priority for ranking microlensing events was calculated based on the return over investment considerations described in \cite{Dominik2010} and filling the estimated available observing time with a fixed sampling time of 1\,hr.

All observations were submitted and monitored automatically by the Target and Observation Manager (TOM) system custom built for this project.  Since that time, our team has developed a general-purpose and formally maintained open-source package for building similar systems called the TOM Toolkit \cite{Street2018_TOM}, based in part on the software used for this program.  Although the majority of the program operated entirely robotically, the TOM system also provided a user interface to enable team members to request additional REA-mode observations if deemed necessary.  

\begin{figure}
    \centering
    \includegraphics{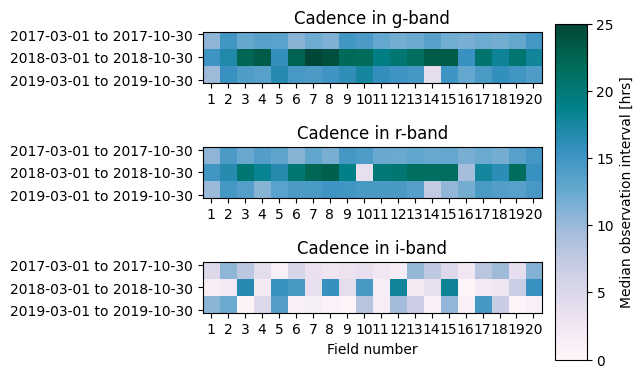}
    \caption{The median interval between sequential observations of each field, plotted as a function of filter and the annual Bulge visibility seasons.}
    \label{fig:real_cadence}
\end{figure}

During the Key Project, LCO undertook a program to re-aluminize the mirrors of its 1\,m and 2\,m telescopes.  This was scheduled to occur at the southern ring sites used in this project between June--November, 2018.  With that in mind, the project focused on acquiring timeseries monitoring observations first, while sequences of multiple long exposures in all filters for all fields were planned for the end of the project, in order to benefit from the re-coated mirrors.  Ideally, such image sets should be acquired on the same night in conditions of good seeing and sky background, for use as deep reference images.  While some deep image sets were acquired in 2019, obtaining extended, more densely-sampled timeseries was prioritized, as this provided better characterization of the brighter events that were our primary targets.  Unfortunately not all field/passband combinations were completed before pandemic-related lock downs interrupted the project's last season (2020A).  Although many of LCO's sites were able to continue operations, the Chilean site which offers the best atmospheric conditions was required to halt operations during the last months of the program by the regional authorities.  

Since the LCO Network is a multi-user facility and not a dedicated survey, it is valuable to compare the cadence realized in practice with the project's original goal of observing each field every $\leq$8\,hrs.  Figure~\ref{fig:real_cadence} presents the median interval between sequential observations for all fields in all three Bulge observing seasons during the project.  This shows that in $i$-band, the median interval is typically much shorter than this, since the majority of REA-mode observations were performed using this filter.  These are intermittant in nature, since they were alert-dependent.  Excluding REA-mode observations, the realized ROME-mode cadence was typically $\sim$10\,hrs $i$ and $\sim$15\,hrs in $g$ and $r$. No change in the project's strategy occurred in 2018, but this coincided with a drop in telescope availability during re-aluminization.  Variable network contention due to changes in other observing programs also affects the execution.  

\section{Data Reduction and Calibration}
\label{sec:pipeline}

The raw data from all LCO images was initially processed by the BANZAI pipeline \cite{BANZAI}, which performs debiasing, flat-fielding etc to remove the instrumental signature.  While this pipeline does extract a source catalog, its aperture-based approach is not ideal for extracting photometry in crowded stellar fields like the Bulge, so the data were subsequently run through the project's own pipeline.  

Difference Image Analysis (DIA) \cite{Alard1998} has become a widely-used approach to reducing crowded imaging data, and the Bramich algorithm \cite{Bramich2008, Bramich2013} has become widely used in the microlensing field (e.g. pyDIA \cite{pyDIA}).  Our team had substantial experience with this algorithm through our existing data reduction pipeline which was developed around the {\texttt DanDIA} library\footnote{http://www.danidl.co.uk/}, written in IDL.  This worked well for reduced small image subsections around a single target, but was prohibitively slow when applied to fullframe images from LCO's 4k$\times$4k Sinistro cameras.  The costs of licensing IDL also motivated us to seek an open-source, Python-based solution.  The intensive computational demands of the algorithm can be mitigated by adapting it to Graphical Processing Units \cite{Hitchcock2021}, but this approach places constraints on the computing hardware required for the pipeline.  Our goal was to develop software that could be run on any CPU, from a laptop to a large cluster.  Two additional factors drove the design of the data handling infrastructure.  Firstly, the multi-site, multi-instrument nature of data from the LCO network, and the large size of the dataset, lends itself to parallelization, since each dataset can be reduced independently, and combined once light curves are extracted.  Secondly, while the data for this work was reduced as a collection at the end of the project, the pipeline was {\em also} run in real-time mode for single-star reductions of the data for specific targets of interest, such as OGLE-2018-BLG-0022 \cite{Street2019}, allowing data to be added to an existing reduction.   

\begin{figure}
\begin{centering}
 \includegraphics[width=1.0\textwidth]{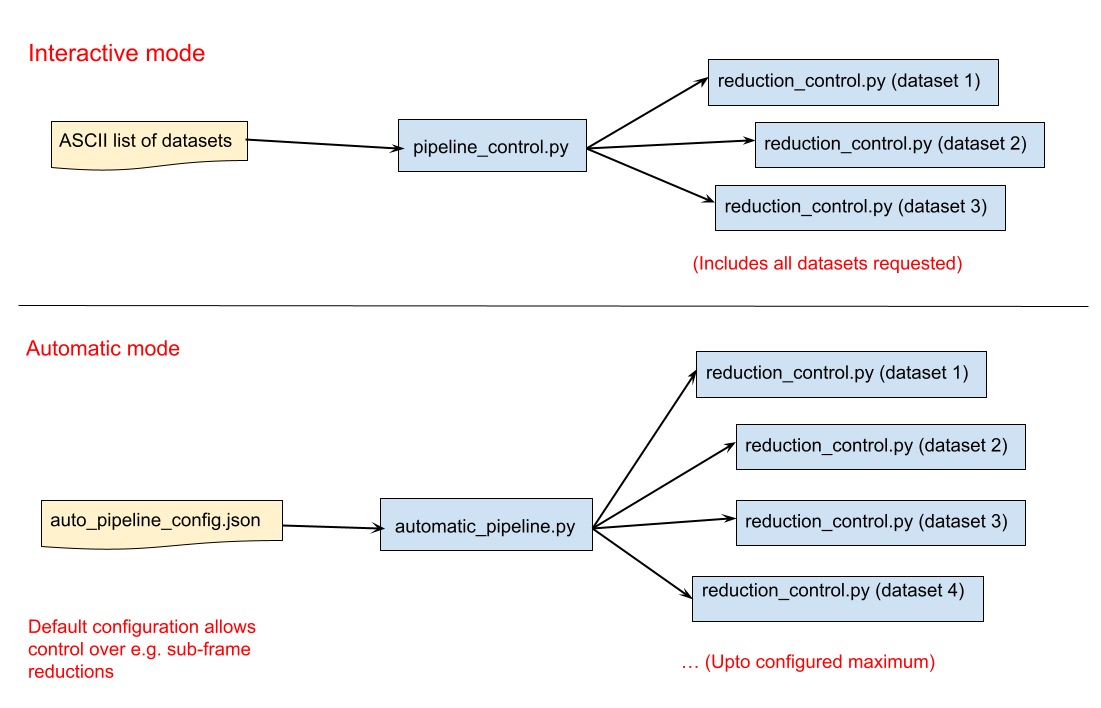} \\
\includegraphics[width=1.0\textwidth]{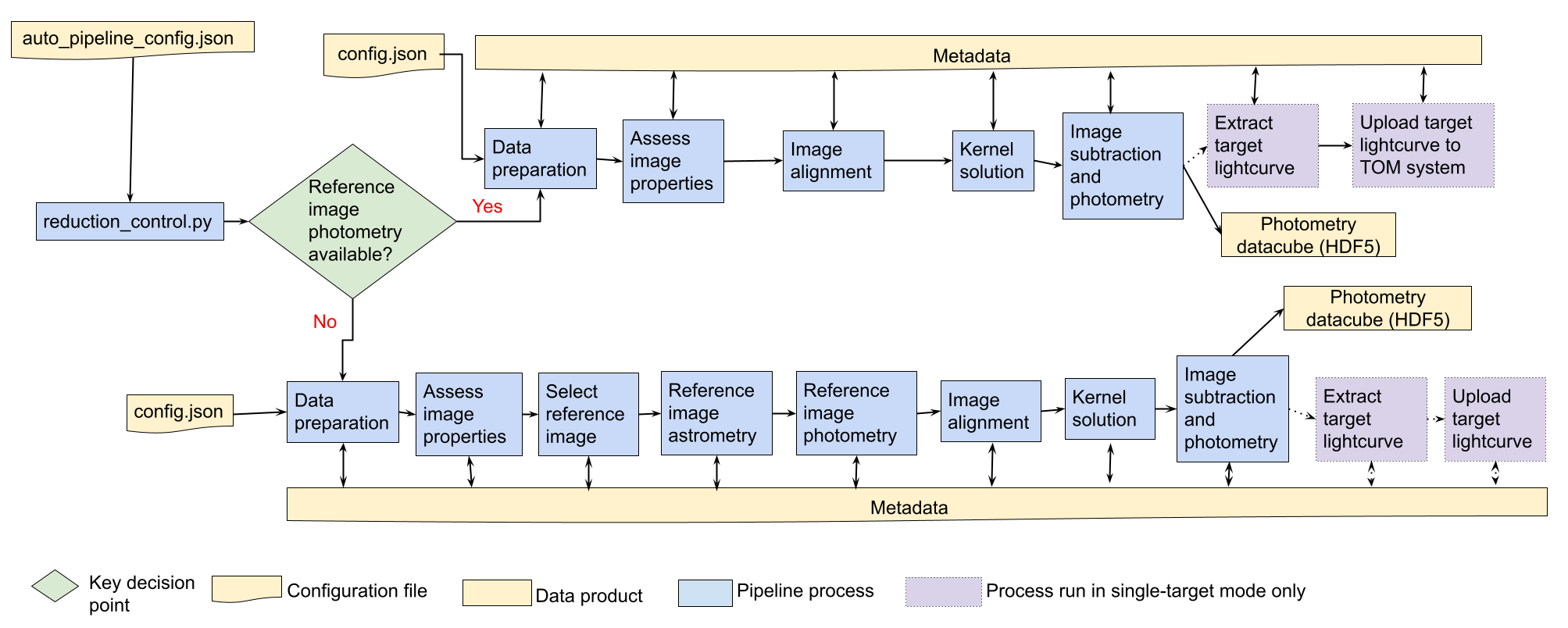}   
\caption{Flowcharts describing pyDANDIA operating in different modes.  (Top) Multiple datasets can be processed in parallel, either using the fully automated pipeline, or in interactive mode.  (Bottom) The alternative workflows that can be taken for the reduction of a given dataset, determined by whether or not a processed reference image is available. \label{fig:pipeline_flowchart}}
\end{centering}
\end{figure}

We developed the pyDANDIA package\footnote{https://github.com/pyDANDIA/pyDANDIA} to provide a Python-based, adaptable data reduction framework capable of reducing fullframe imaging data from multiple instruments in a highly automated manner.  The pipeline is structured into the following stages, which are graphically illustrated in Figure~\ref{fig:pipeline_flowchart}.  

\subsection{Data structure, configuration and preparation}
For the purposes of reduction, the data were grouped into \texttt {datasets} according to combinations of survey \texttt{field} pointing, observing site, enclosure, telescope, instrument and filter.  The naming structure adopted enables the pipeline to uniquely distinguish datasets when some LCO facilities have multiple enclosures per site, multiple telescopes per enclosure, and multiple instruments per telescope.  

Instrument-signature corrected image data products acquired by the Key Project were downloaded automatically from the LCO Data Archive\footnote{https://archive.lco.global} by the project's data handling infrastructure as soon as they were available.  pyDANDIA was designed to operate together with this architecture, but also to operate as a stand-alone pipeline to facilitate its use in other contexts. 

After sorting, the pipeline process \texttt{reduction\_control.py} is designed to run an automated, end-to-end reduction of a single dataset, and pyDANDIA includes control software \texttt{pipeline\_control.py} designed to manage multiple parallel instances of the pipeline, run manually in an interactive mode.  This enabled us to parallelize the reduction of different datasets on different CPUs in a multi-processor computing cluster.  \texttt{pipeline\_control.py} offers the user more control over human-monitored reductions, and can parallelize the end-to-end reduction, or individual sections of the pipeline.  In addition, pyDANDIA provides \texttt{automatic\_pipeline.py}.  This program is designed to run the pipeline in fully automated operation, which is useful for real-time, ``quick-look'' reductions.   All modes of the pipeline can be configured to process subframes around a specific target, although fullframe mode was used to reduce the data for this project.  

The pipeline is configured by means of a set of files in JSON format.  The characteristics of each instrument, such as its gain and readout noise, are defined in a dedicated configuration file, while parameters governing the DIA process are provided in the file \texttt{config.json}.  A third file, \texttt{auto\_pipeline\_config.json} provides control over the directory structure and number of simultaneous processes allowed for automated, parallel reductions.  

The first stage of the pipeline reviews the data available in a given dataset and loads the necessary configuration files for the instrument.  All of the metadata relating to a single dataset is stored in a single \texttt{metadata} file.  This is a multi-extension FITS binary table file that is used as the single point of reference throughout the rest of the pipeline, allowing relevant information to be passed between different stages.  In a completed reduction this file includes tables describing:

\begin{itemize}
    \item the data architecture, 
    \item the configurable parameters used for the reduction, 
    \item a summary of essential information from the image headers, including timestamp information,
    \item a status table recording the which stages of the pipeline have been performed for each image,
    \item basic statistics calculated for each image,
    \item the pixel coordinates of the image stamps used for later sections of the pipeline,
    \item the dimensions of the PSF,
    \item a ranking of all images evaluating their quality as a reference image,
    \item photometry and astrometry of stars detected in the reference image for the dataset, 
    \item the parameters of the photometric calibration of the reference image to the photometric catalog,
    \item a table recording the versions of the pipeline software used for key stages of the reduction.
\end{itemize}

Since the pipeline is designed to operate in real-time as well as `offline' process modes, data can be added to a pre-existing reduction.  This is managed by each stage of the pipeline refering to the metadata's status table to identify only those frames which have not yet been reduced by the current stage.  The modular design of the pipeline allows the user to run all functions separately if desired. 

The data preparation stage computes the pixel dimensions of image \texttt{stamps}, sub-dividing the fullframe image into a configurable number of sections.  This sectioning enables later stages of the pipeline to be  optimized for greater efficiency.  In real-time mode, the pipeline is typically configured to process just a single stamp in the center of the frame, since this mode is normally used to process the data for a single object.  For the ROME survey processing, 16 stamps were used, covering the fullframe images.  The stamps are typically $\sim$1000$\times$1000\,pix.  

Care was taken to mask out bad pixels, dead columns, saturated stars and other artifacts that can strongly distort subsequent image resampling stages.  The pipeline builds on the Bad Pixel Mask (BPM) produced by the BANZAI pipeline, and adds to this masks for those pixels that are saturated in each image.  The column bleeds of severely saturated stars are masked using the \texttt{binary\_dilation} function from the \texttt{scipy.ndimage.morphology} library, and the code also checks for negative pixel values. A BPM is initially appended to each data image as an extra FITS image extension, for use in later stages. 

The data preparation stages of the pipeline also perform a preliminary object detection in each image in order to estimate the number of stars in the frame, the mean PSF Full Width Half Maximum (hereafter FWHM), and the sky background.  These parameters, together with information on telescope pointing telemetry collated from the image headers, is then used to perform a quality assessment for each image.  Image flagged by this stage are not reduced further.  

\subsection{Reference image selection and analysis}
pyDANDIA includes a function to automatically select the best-available image from a dataset to use as a reference.  This function allows the reduction to be fully automated, and is particularly valuable in real-time mode.  Since the functionality is provided, we describe the procedure below, but note that for the ROME Data Release this selection was overridden (see below).  

All images in the dataset are ranked based on the FWHM, the Moon phase (if that information is available in the image FITS headers), noise contribution of the sky background and the number of stars detected in the image.  The selection threshold applied for the FWHM can be configured via the JSON files.  

If no available image meets these criteria, rather than ranking images based on sharpness and applying a cutoff to the sky background, the aim was to maximize the expected signal-to-noise ratio of a typical target. Seeing entered the equation through the number of pixels over which the signal was distributed.
Assuming a typical target magnitude, the ranking of selected images is based on the contribution of the sky background and readout noise to the total noise budget for the image.  Seeing enters the equation through the number of pixels.  

If any images are selected by this process, the highest ranking one is selected as the reference and copied to a subdirectory.  This image is then used as the photometric reference for the remainder of the reduction. A goal for the future development of the pipeline is to support the co-addition of multiple images to provide a deeper reference image.  Since complete deep image sets for all twenty fields in all three filters were not obtained before the end of the Key Project, single-image references were used for this data release. 

As one of the goals of the ROME survey was to provide color information for all stars, it was necessary to coordinate the reference images selected for datasets from a given telescope taken in different filters.  We selected `triplets' of reference images, taken on the same night from the same camera, for the telescopes used for the ROME strategy, ensuring that the reference image photometry from those datasets could later be used for color analysis.  pyDANDIA includes tools for identifying such triplets, and allows the user to override the automatic choice of reference image accordingly. 

Once a reference image is assigned for a dataset, the pipeline performs object detection, and the resulting \texttt{starcatalog} is appended to the metadata.  A subset of stars from the center of the frame was cross-matched against the {\em Gaia} Data Release 2 (DR2) sources within $\sim$30\,arcmin of the nominal field center for all ROME fields, and a six-parameter transformation calculated in order to derive astrometry for all detected stars. {\em Gaia} DR2 was used because it was available early on in the ROME survey and includes static astrometry that is sufficient for our purposes, since the proper motions of stars in the Bulge is relatively small.  Once a satisfactory fit was achieved, the pipeline cross-matched the fullframe catalog of detected sources against both the {\em Gaia}-DR2 source list, as well as that from the VPHAS$+$ survey \cite{Drew2014} (Vega system photometry). The latter survey provides SDSS-$u, g, r$ and $i$-band photometry which was used as a basis for the photometric calibration.  

PSF-fitting photometry was then performed on the reference image for each dataset, using pyDANDIA's built-in functions. 
A selection of stars to use to model the image PSF was made automatically from objects detected in the center of the image.  This criterion was introduced purely because of the ample number of stars available these fields, to improve the computation time of this stage; in more sparsely-populated fields, this selection would be removed.  Stars with close neighbors with a flux ratio higher than a configurable threashold were excluded, to avoid heavily blended objects.  
The image data for the selected stars is then combined to build a PSF, using an iterative procedure that fits the first-pass PSF to objects detected in the wings of PSF stars to subtract these from the image data before rebuilding the final PSF from the `cleaned' PSF stars.  A number of PSF functions were trialed, and a 2D Moffat function was found to fit the data well most consistently.  

Stars with both ROME and VPHAS$+$ measurements allowed us to derive an approximate, two-parameter linear function which was used to transform the instrumental photometry for the reference image of each dataset ($m_{inst}$) in passband $f$ to the VPHAS$+$ system ($m_{cal}$).  

\begin{equation}
    m_{cal}(f) = a_{0}(f)*m_{inst}(f) + a_{1}(f),
    \label{eqn:phot_cal1}
\end{equation}

where $a_{0}$, $a_{1}$ are fitted co-efficients.  Although these steps are designed to run automatically in the real-time mode, the astrometric and photometric calibrations for all datasets were reviewed manually for the ROME survey.  The VPHAS$+$ survey footprint overlaps that of ROME almost entirely, allowing this procedure to be used for all fields except ROME-FIELD-20.  In this case, reference images for each dataset were selected from the same night as those used for the nearby ROME-FIELD-19.  The photometric calibration coefficients derived for those datasets for ROME-FIELD-19 were then applied to the corresponding datasets for ROME-FIELD-20.  

\subsection{Image alignment, subtraction and timeseries photometry}
Before Difference Image Analysis can be performed, it is first necessary to geometrically register all images in the dataset with the reference image. pyDaDIA uses the \texttt{phase\_cross\_correlation} function from \texttt{scikit-image}'s registration library to derive initial x,y pixel offsets for all images. Using this as a starting point, the pipeline then determines a full matrix transformation including shifts, rotation and scaling.  

Variations in atmospheric transparency, seeing variations, and different exposure times are effectively handled by constructing a convolution kernel that ``smoothes'' a reference image to produce an optimal difference image in a least squares sense.  There is no need for comparison stars, as is common in differentil photometry, because the photometric scale factor captures the variations caused, for example, by changes in atmospheric transparency and exposure time.  In this sense, the whole field acts as a comparison star \cite{Bramich2015}.  Different methods are used in order to find the best kernel. In this work, we have used a simplified numerical kernel as introduced by \cite{Bramich2008}. The kernel solution includes an estimate of the background but for numerical stability and to capture background gradients, the background is first subtracted at the subimage level. The model image is obtained from reference image and 2D background model: 
\begin{equation}
{\rm Model} = {\rm Kernel} \otimes {\rm Reference} + {\rm Background}.
    \label{eqn:kernel}
\end{equation}

Solving for the kernel solution requires constructing a computationally expensive design matrix of the least squares problem, analogous to finding the slope and intercept of linear regression. In addition, the computation time scales with the kernel width to the power of four. A kernel width twice as large requires 16 times more computational effort. Instead of repeatedly computing the design matrix for each reference image, the noise model and the bad pixel mask are kept constant. The convolution kernel for each image can be estimated consistently and quickly from the image. When keeping the design matrix fixed, the noise model is also assumed to be fixed.  When the design matrix is calculated for each image, the square root of the reference image is used as the inital noise model, and the model images is contructed by convolving the first kernel solution with the reference image, followed by repeating the design matrix construction for the revised model image.  The approach requires careful alignment of the images with subpixel accuracy, which is done as part of the image alignment stage, which first finds the shift with respect to the reference image and then resamples the image as an affine transformation using the RANSAC algorithm \footnote{as implemented in https://scikit-learn.org/}. 

PSF fitting is then used to perform photometry on the subtracted images for all stars in the catalog for each dataset, and the resulting timeseries photometry undergoes a post-processing step to evaluate the quality of each photometric measurement.  An integer \texttt{qc\_flag} parameter is assigned to all timeseries photometry points. By default, good quality measurements receive \texttt{qc\_flag}=0, while bitmask values are added to this flag to indicate different data issues (this is described in more detail in Section~\ref{sec:data_products}).  The thresholds used for all quality control assessments can be configured by the user. 

The procedure above outputs timeseries photometry for all stars in the field of view, calibrated to VPHAS$+$, for each dataset (site-telescope-instrument-filter combination) separately, as a datacube stored in Heterogeneous Data Format 5 (HDF5) format.  

It is possible to configure the reduction of a dataset in `single-target mode'.  This is useful for real-time reductions of data obtained during follow-up observations of a specific transient alert for example.  In this mode, the pipeline is made aware of the coordinates of the target object, and extracts the timeseries photometry for that object in CSV format at the end of the reduction.  pyDANDIA includes a module which enables the automatic upload of the target lightcurve to a Target and Observation Manager (TOM) system (also known as Marshals).  Such systems are used to automate the observing programs for a number of major projects, including the microlensing programs at Las Cumbres Observatory, so pyDANDIA has been designed to integrate with LCO's TOM Toolkit package \cite{Street2018_TOM}. 

\subsection{Field data products and dataset normalization}
Since the timeseries from each site is subject to diurnal gaps, it is valuable to be able to combine the photometry of a given field pointing in a given filter from all instruments, to achieve 24-hr coverage.  We refer to the resulting data as the field data products.  

The first step towards building the field data products is to crossmatch the source catalogs detected in the reference images for all datasets obtained for a given field.  We nominated the datasets obtained from Chile, Dome A, telescope 1m0-05 and camera fa15 as the `primary reference' datasets ($g,r,i$) for all fields, since the conditions at this site consistently have the best seeing and transparency of the LCO network.  Objects detected in the primary reference datasets in all passbands were combined to form a single source catalog for each field.  The source catalogs of the other datasets were crossmatched against the field catalog, with stars being added if they were detected in some, but not all, datasets.  This occurs as a natural consequence of small errors in telescope pointing between the facilities.  The combined field catalog was then crossmatched against the {\em Gaia}-EDR3 catalog for all 20 fields.  

This process creates the `crossmatch table' for each field in the survey, a multi-extension FITS binary table which is used to store metadata relevant to the collection of datasets, as well as a reference `field index' used to locate the photometric array entries of specific stars and images.  Each star in the field catalog is assigned a unique field identifier for future reference. 

The crossmatch table is then used to combine the timeseries photometry from all datasets (in all passbands).  However, as a single field typically contains over 300,000 stars and $>$2000 images, the combined photometry can total $>$100\,GB, making it unwieldy to store in a single file.  Instead, each field is subdivided into four equal quadrants and all stars in the field catalog are assigned to a quadrant based on their coordinates. The timeseries photometry for each quadrant is stored as a separate HDF5 file.  Tools are provided in pyDANDIA for handling these data products. 

At this stage of processing, the full 3-yr lightcurve for any star in a field can be extracted, using either the instrumental or calibrated photometry. However, although the photometry for each dataset was calibrated relative to the VPHAS$+$ catalog, there remain small offsets between the calibrated timeseries photometry for different datasets. 

Two factors contribute to these offsets.  The first is due to the fact that the reference images for each dataset were taken by different facilities on different nights.  This can be measured by comparing the calibrated photometry from the reference images for different datasets in the same file.  Variable stars were excluded by selecting only those showing relatively low photometric residuals, and then a two-parameter transformation between the photometry in each dataset ($d$), $m_{corr}(d, f)$, and that of the primary reference dataset in the corresponding passband, $m_{pri}(f)$, was calculated, 

\begin{equation}
    m_{corr}(d,f) = b_{0}(d,f)*m_{cal}(d,f) + b_{1}(d,f),
    \label{eqn:phot_corr}
\end{equation}

where $b_{0}$, $b_{1}$ are the fitted coefficients. 
The resulting function was used to normalize all datasets to the primary reference dataset in each filter, meaning that the primary reference dataset can be used to derive color information for all stars, noting that this calibration process does not account for extinction within the Bulge fields.  

The second factor is that the photometry for a given star in different datasets also depends on the quality of its PSF fit in the reference images of each dataset.  This can only be evaluated for each star in all datasets, and the offsets from the primary reference datasets in each case were calculated by binning the photometry from each dataset into bins of 1-d in width, and evaluating the residual of the primary reference binned photometry minus that of the binned datasets.   Some stars were not measured in the Chilean primary reference datasets due to telescope pointing offsets.  In these cases, the data from South African Dome A, followed by the Australian Dome A, were used (in order of preference due to site conditions) as the primary reference.  A small fraction of stars were only measured in other datasets, and in these cases no second factor normalization was applied.  

Once both normalizations are calculated and applied for all lightcurves, the field data products store instrumental, calibrated and normalized photometry, together with the normalization coefficients, so that users can select their preferred data.  

\section{Results}
\label{sec:results}
\begin{figure}
\begin{centering}
\begin{tabular}{cc}
 \includegraphics[width=0.5\textwidth]{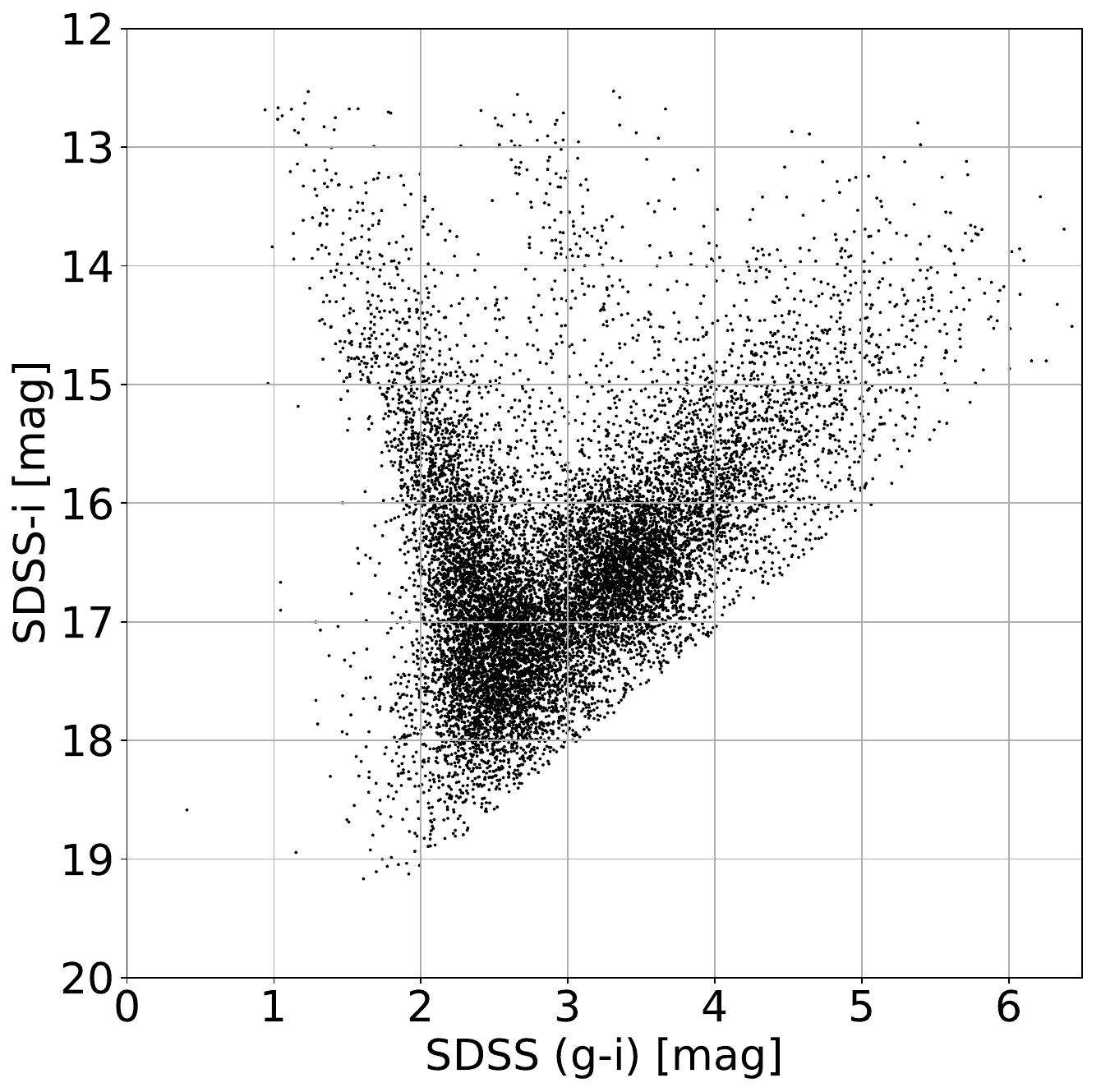} &
\includegraphics[width=0.5\textwidth]{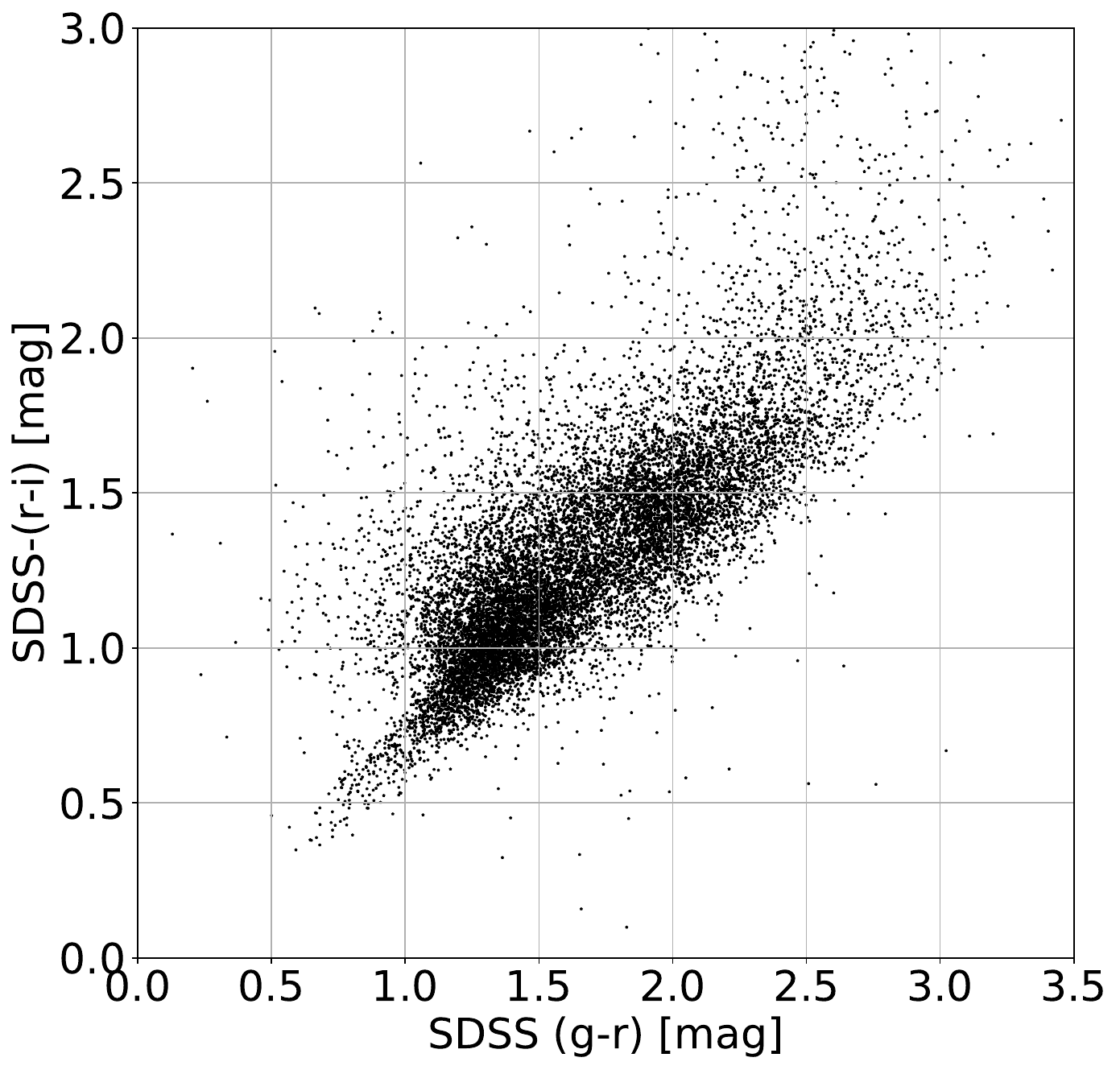} \\  
\end{tabular}
\caption{(Left) color-magnitude and (right) color-color diagrams for ROME-FIELD-01, plotting every 5th datapoint to reduce the plot file size.  No corrections have been made for extinction or reddening, which is highly spatially variable within this field. \label{fig:ROME01_CMD}}
\end{centering}
\end{figure}

\begin{figure}
\begin{centering}
\includegraphics[width=0.5\textwidth]{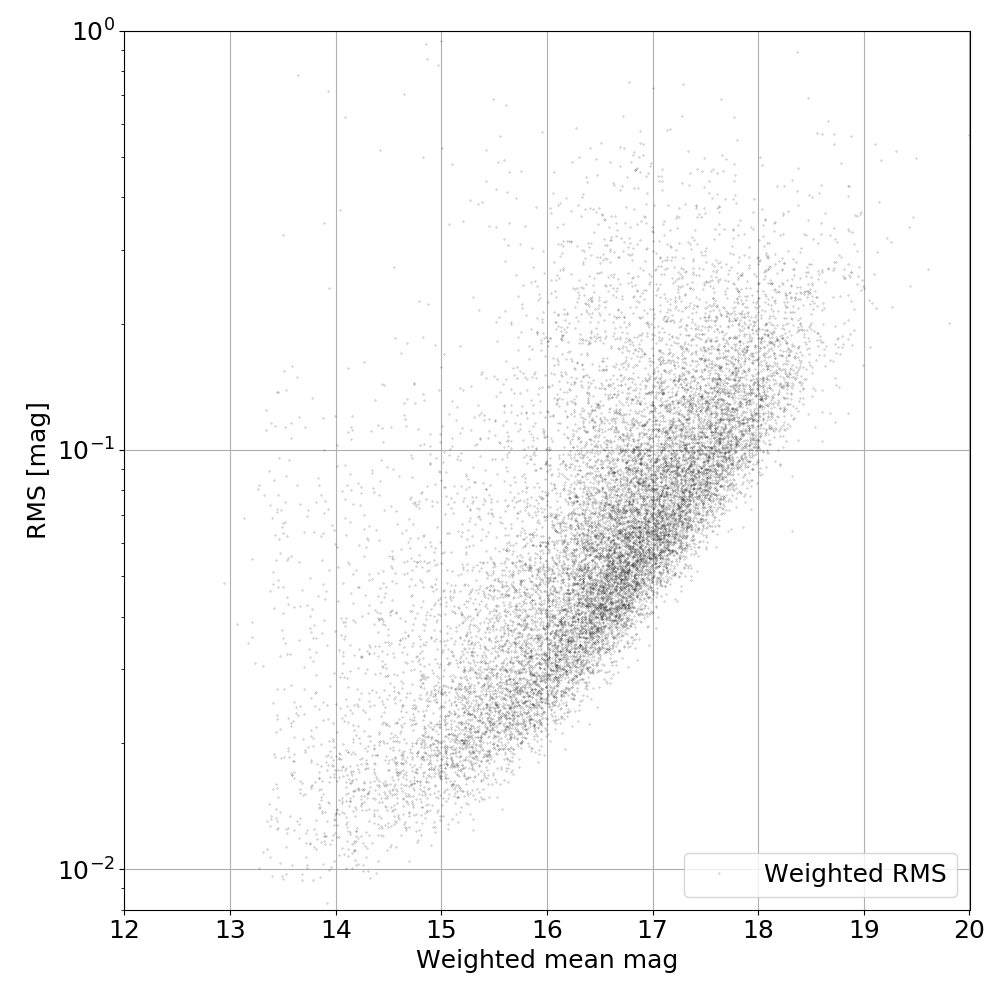} 
\caption{Weighted root-mean-square magnitude deviation of the timeseries photometry in SDSS-$i$ band as a function of weighted mean magnitude for stars in quadrant 1 of ROME-FIELD-01. \label{fig:ROME01_RMS}}
\end{centering}
\end{figure}

In this section, we explore the data for ROME-FIELD-01, as an example of the data products included in this release.  
Figure~\ref{fig:ROME01_CMD} highlights the multi-band photometry  generated for each field.  These plots include photometry for a random sample of 1/5 of the stars that were measured in all three passbands with a precision of $\leq$0.1\,mag.  This field was chosen as an example of the complex stellar environment in these Galactic Bulge fields, since it includes both dust clouds and highly variable extinction and a foreground open cluster, NGC~6451 (see Figure~\ref{fig:survey_map}).  These features are reflected in the `tail' of bright and blue stars in the color-magnitude diagram as well as the wide range of $(g-i)$ colors. Figure~\ref{fig:ROME01_RMS} illustrates the photometric precision of the timeseries data, giving the example of data from quadrant 1 of ROME-FIELD-01.  The photometric uncertainties were used as inverse variance weights for the root-mean-square and mean magnitude estimates, to minimize the impact of outliers, and points flagged as poor quality by the pipeline were excluded. 

\begin{figure}
    \centering
    \begin{tabular}{cc}
        \includegraphics[width=0.4\textwidth]{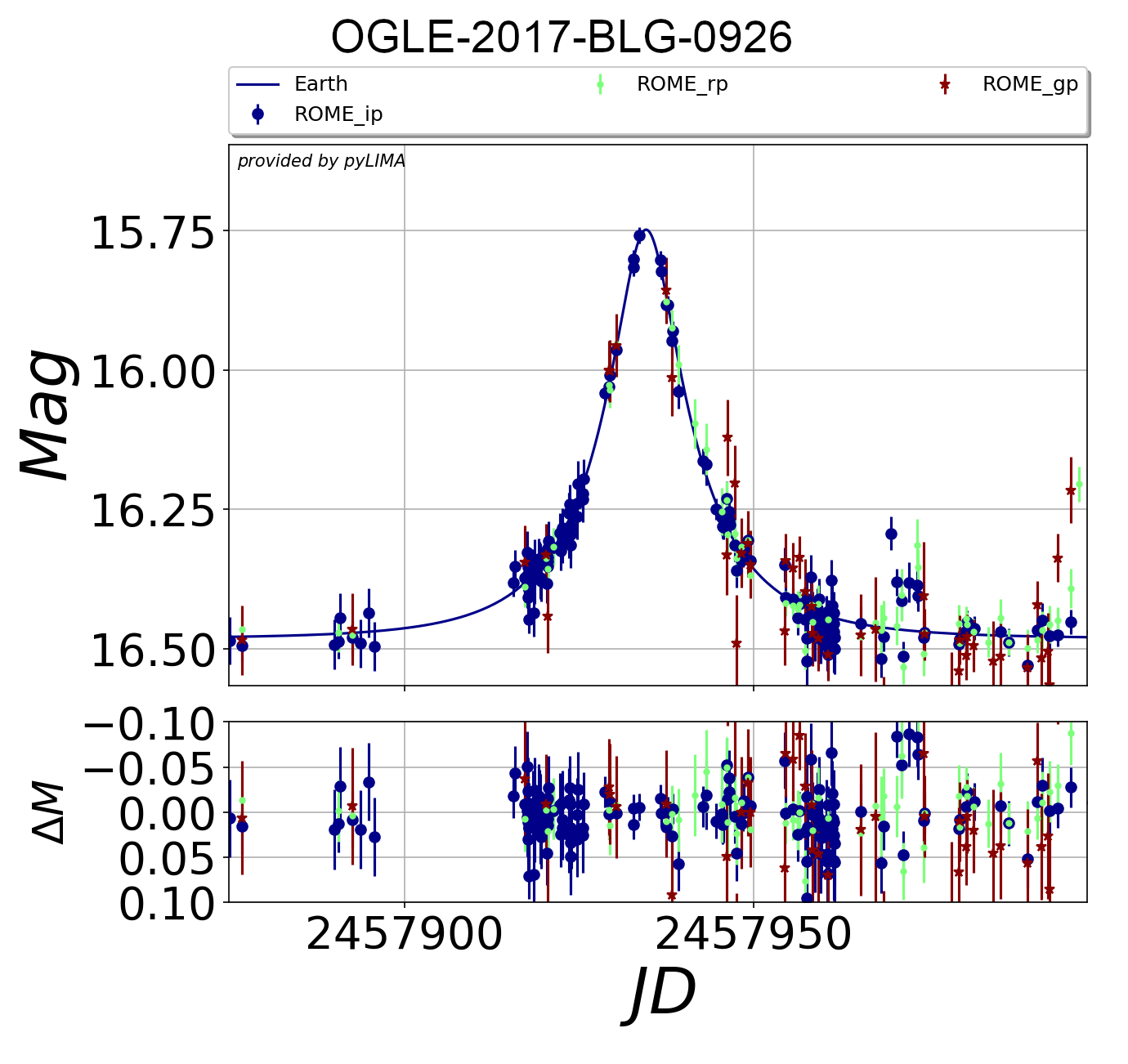} & 
        \includegraphics[width=0.4\textwidth]{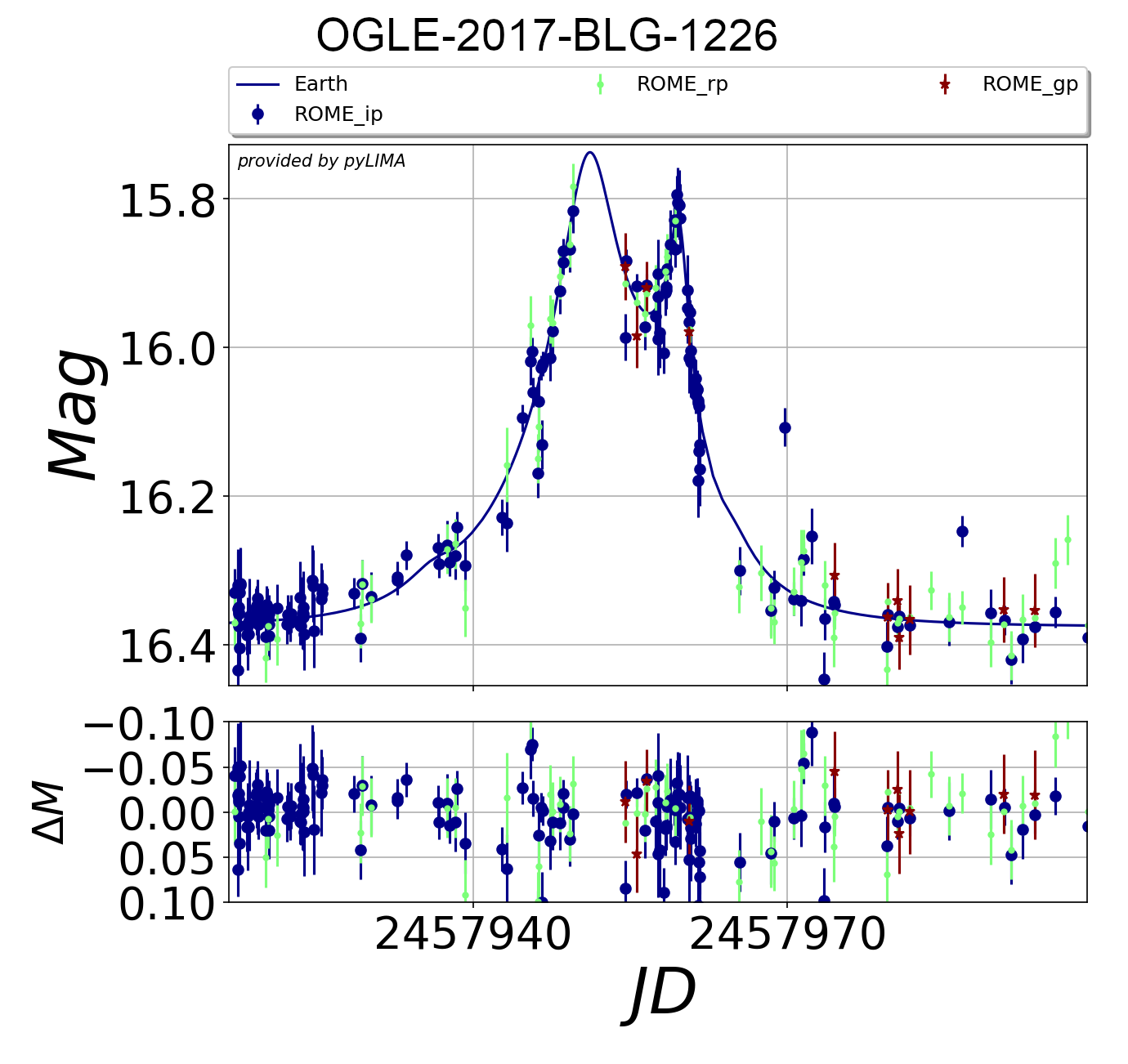} \\
        \includegraphics[width=0.4\textwidth]{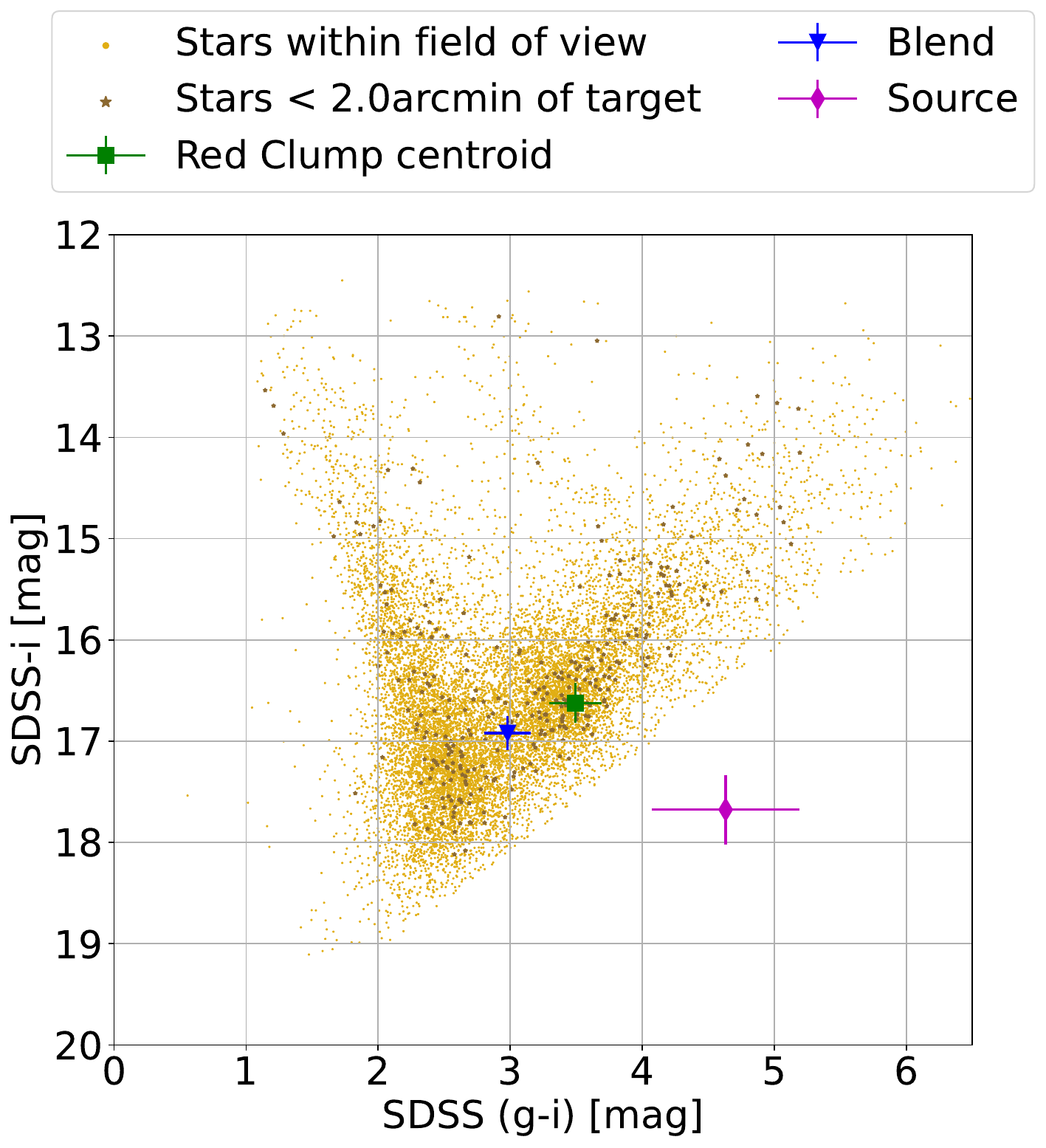} &
        \includegraphics[width=0.4\textwidth]{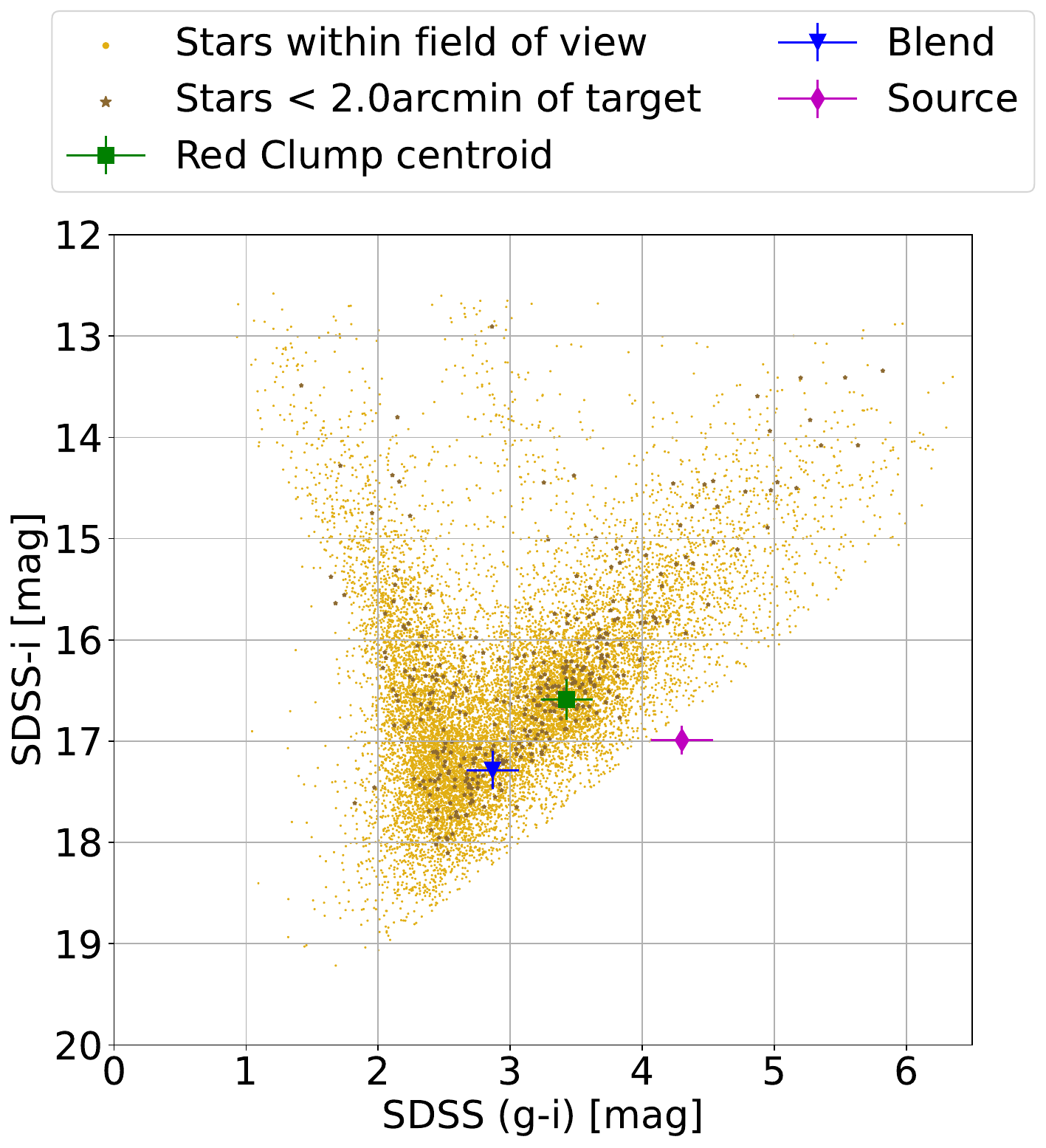} \\
        \includegraphics[width=0.4\textwidth]{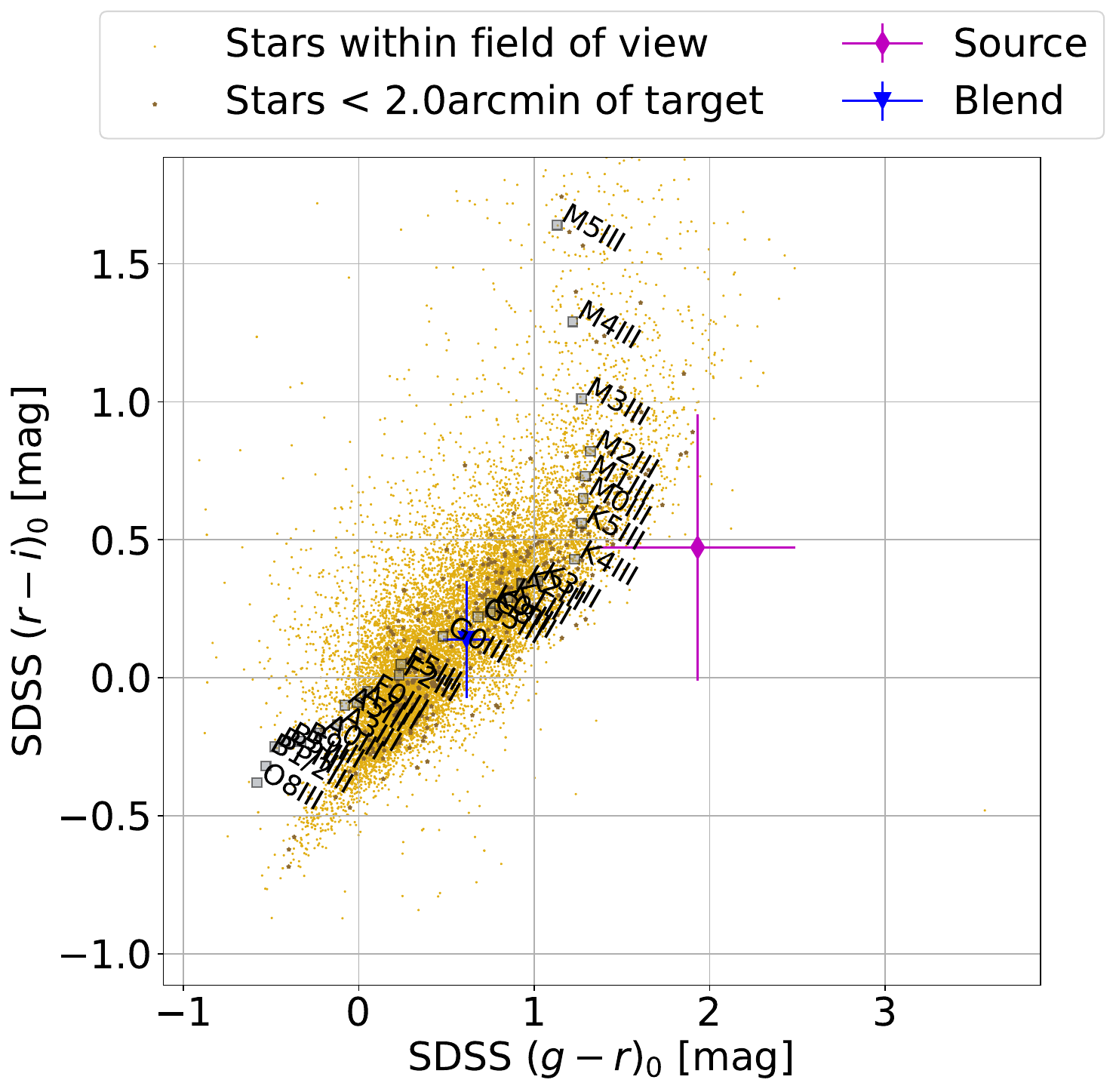} &
        \includegraphics[width=0.4\textwidth]{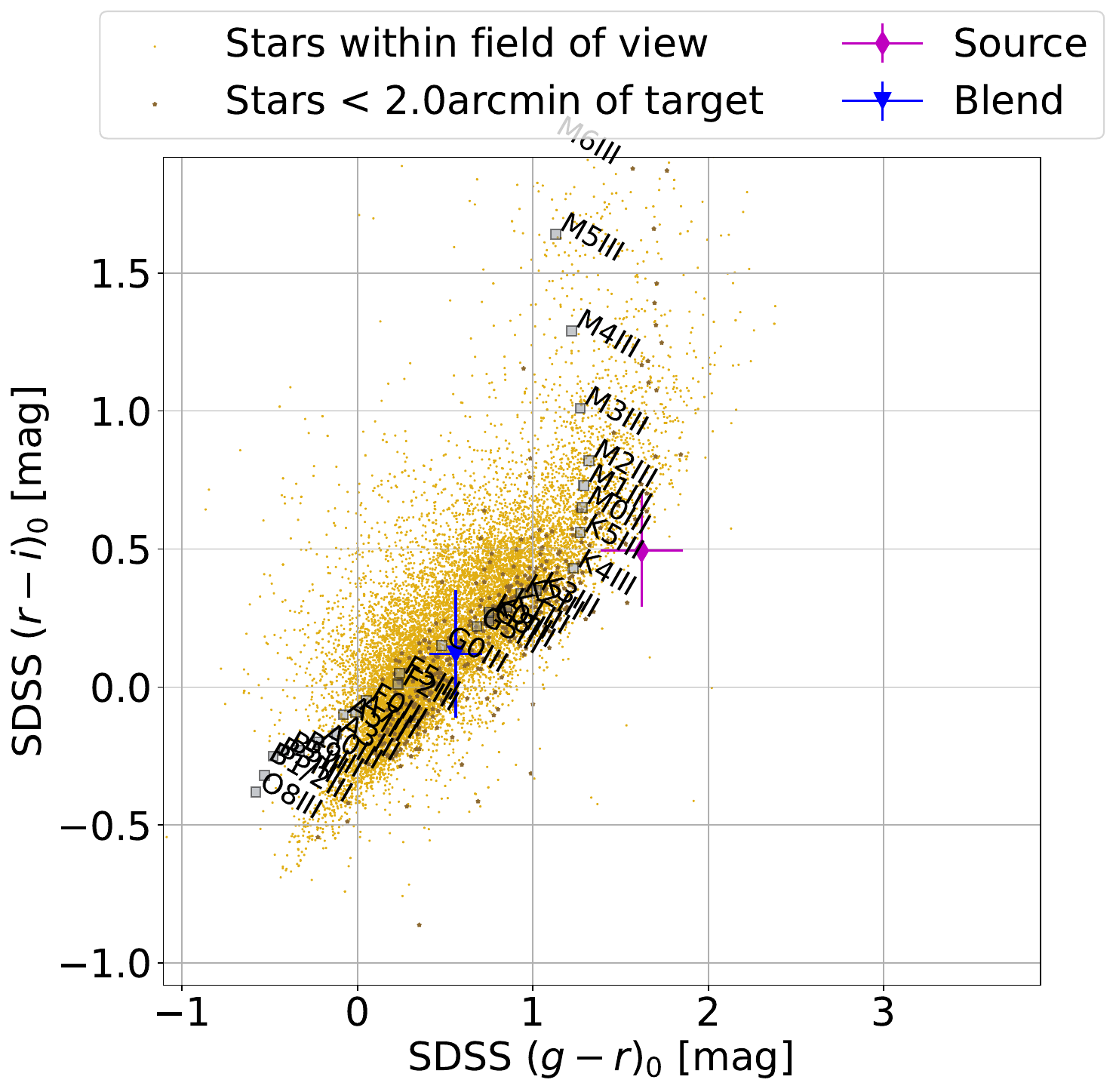} \\
    \end{tabular}
    \caption{(Top) Light curves (zoomed in around the peak of the event) for examples of microlensing events found within ROME-FIELD-01, (middle, bottom) the color-magnitude and color-color plots for the regions including each event, with the fluxes of the lensed source and blended stars indicated. The color-color diagrams have been corrected for extinction and  reddening based on the measured position of the Red Clump. Data for every 5th star in the background field has been plotted to reduced the plot file size.  }
    \label{fig:mulens_star_lcs}
\end{figure}

The catalog of objects detected in ROME data was crossmatched by position against public lists of events detected by the OGLE, MOA, KMTNet and Spitzer microlensing surveys.  This process identified a total of over 1,100 microlensing events alerted within the ROME footprint during the 3-yr survey period.  Full analyses of these events are the subject of independent papers, e.g. \cite{Street2019}, but examples of data on two example events from the ROME-FIELD-01 field are presented in Figure~\ref{fig:mulens_star_lcs}.  The pyLIMA modeling software \cite{pyLIMA} was used to fit point-source, point-lens (PSPL) or uniform source, binary lens (USBL) models if the morphology of the light curve indicated binarity.  
\setcounter{footnote}{0} 
These models are parameterized as follows: $t_{0}$ indicates the time of the event peak, $u_{0}$ is the impact parameter, $t_{\rm{E}}$ is the Einstein crossing time, $\rho$ describes the angular size of the source star in units of the angular Einstein radius, $\theta_{\rm{E}}$.  For binary models, $q$ represents the mass ratio of the binary components, $s$ describes their angular separation in units of the $\theta_{\rm{E}}$, and $\alpha$ is the angle of the source's trajectory relative to the axis of the binary lens.  The parameters of the best-fitting models for each event are given in Table~\ref{tab:model_parameters}.  Though an exhaustive search of parameter space for definitive models of all events within the ROME survey is beyond the scope of the current paper, these preliminary models are consistent with independent results from the RTModel Real-Time Modeling system\footnote{https://www.fisica.unisa.it/GravitationAstrophysics/RTModel.htm} \cite{Bozza2010}, which analyzed data for these events taken by other observatories.  Due to the dense crowding in the Galactic Plane, the lensed flux from the source stars ($f_{s}$) of microlensing events is almost always blended with flux from neighboring, unlensed objects ($f_{b}$).  Since the source flux is measured at a range of different magnifications during the event, the microlensing models can be used to infer the unlensed source and blend flux, as $f(t) = f_{s} A(t) + f_{b}$.  $f_{s}$ and $f_{b}$ are typically derived from the model fit as additional parameters for each lightcurve. When these parameters are derived for the $g$, $r$, $i$ lightcurves, the source and blend magnitudes can be placed on the color-magnitude and color-color diagrams for the field.  As is routine for the analysis of microlensing events in the Galactic Bulge, the well-defined color and magnitude of Red Clump giants ($M_{g,RC,0} = 1.331\pm0.056$\,mag, $M_{r,RC,0}$ = 0.552$\pm$0.026\,mag, $M_{i,RC,0} = 0.262\pm0.032$\,mag, \cite{Ruiz-Dern2018}) can be used to determine the extinction and reddening towards the event.  Accounting for the highly variable extinction in Bulge fields, it is common to make this measurements using only stars within a $\sim$2\,arcmin radius of the event.  The extinction-corrected fluxes in the three bands can then be used to constrain the spectral type of the source star, and providing an essential constraint on its angular diameter and distance.  This strategy is widely used in the analysis of microlensing events, and allows the angular Einstein radius, $\theta_{\rm{E}}$, to be measured and hence the lens mass.  Street et al.~(2019) \cite{Street2019} describes in detail how the ROME data can be used for this purpose.  

\begin{table}[]
    \centering
    \begin{tabular}{|c|c|c|c|}
    \hline
      Parameter & Units & OGLE-2017-BLG-0926 & OGLE-2017-BLG-1226\\
    \hline
      $t_{0}$   & [HJD] days  & 2457934.629$\pm$0.131 & 2457951.487$\pm$0.088 \\
      $u_{0}$   &   & 0.263$\pm$0.069 & 0.420$\pm$0.0465 \\
      $t_{\rm{E}}$ & days & 16.601$\pm$2.677 & 10.009$\pm$0.571 \\
      $\rho$    &       & - & 0.0$\pm$18.7 \\
      $\log_{10}(q)$ &       & - & -0.301$\pm$0.025\\
      $\log_{10}(s)$ &       & - & -0.223$\pm$0.007\\
      $\alpha$ & rads   &   & 4.49$\pm$0.024 \\
      $\chi^{2}$ &    & 4388.56 & 1487.24\\
      N datapoints &   & 1334 & 1230 \\
    \hline
    \end{tabular}
    \caption{Microlens model parameters for events OGLE-2017-BLG-0926 and OGLE-2017-BLG-1226}
    \label{tab:model_parameters}
\end{table}

\begin{figure}
\begin{centering}
\begin{tabular}{cc}
\includegraphics[width=0.45\textwidth]{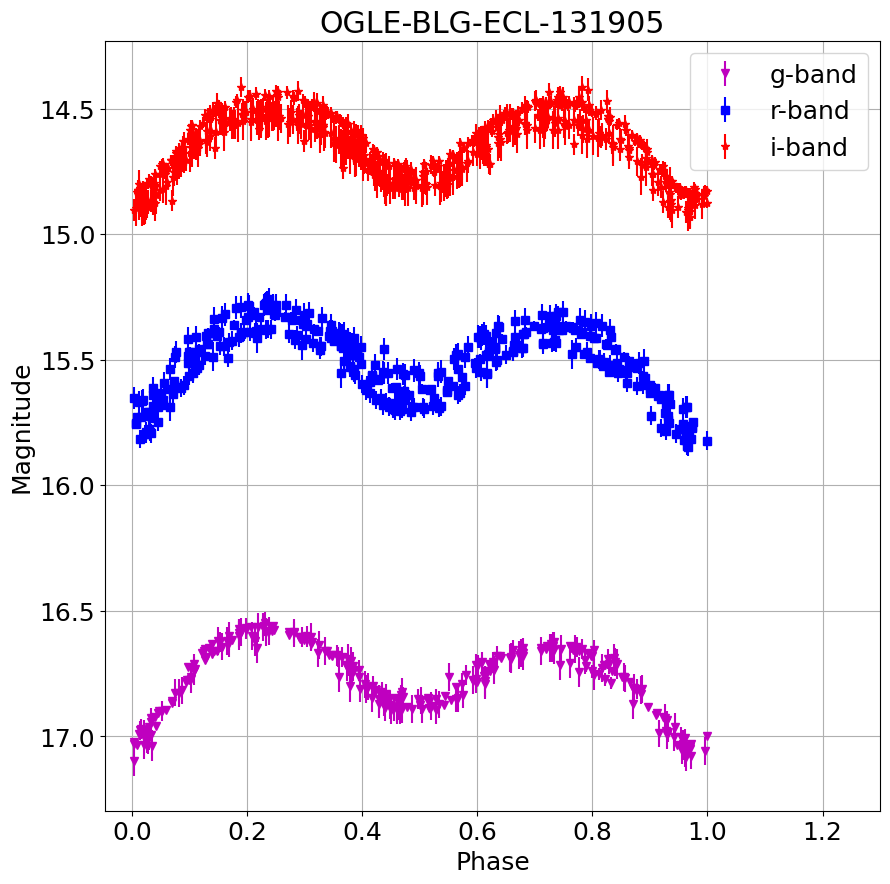} & 
\includegraphics[width=0.45\textwidth]{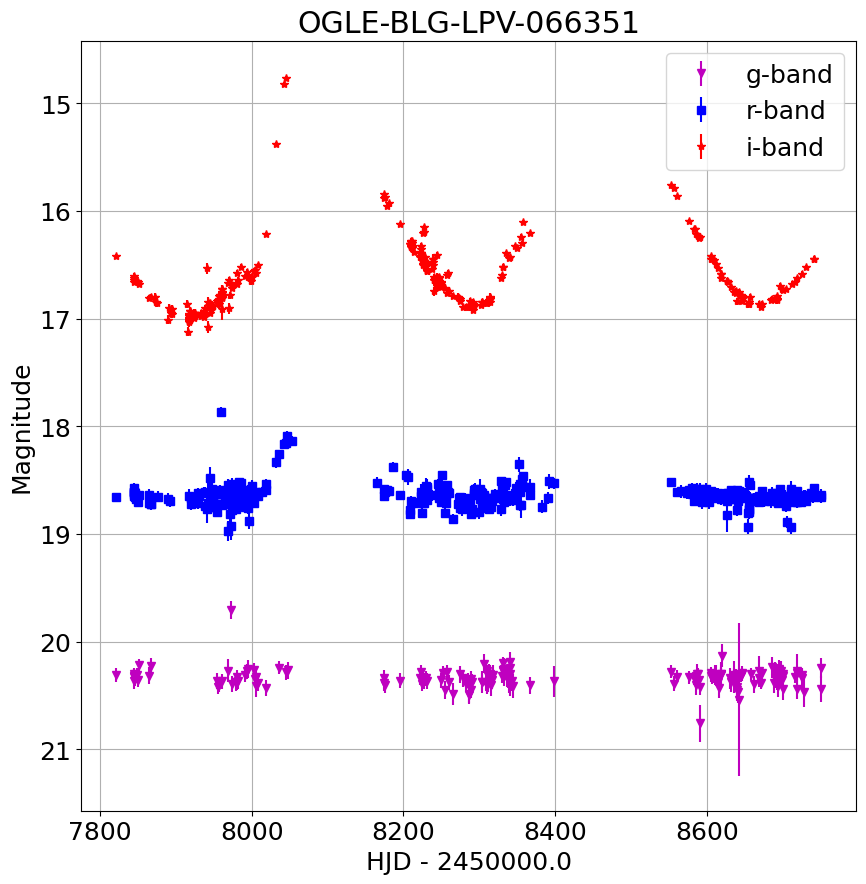}\\
\includegraphics[width=0.45\textwidth]{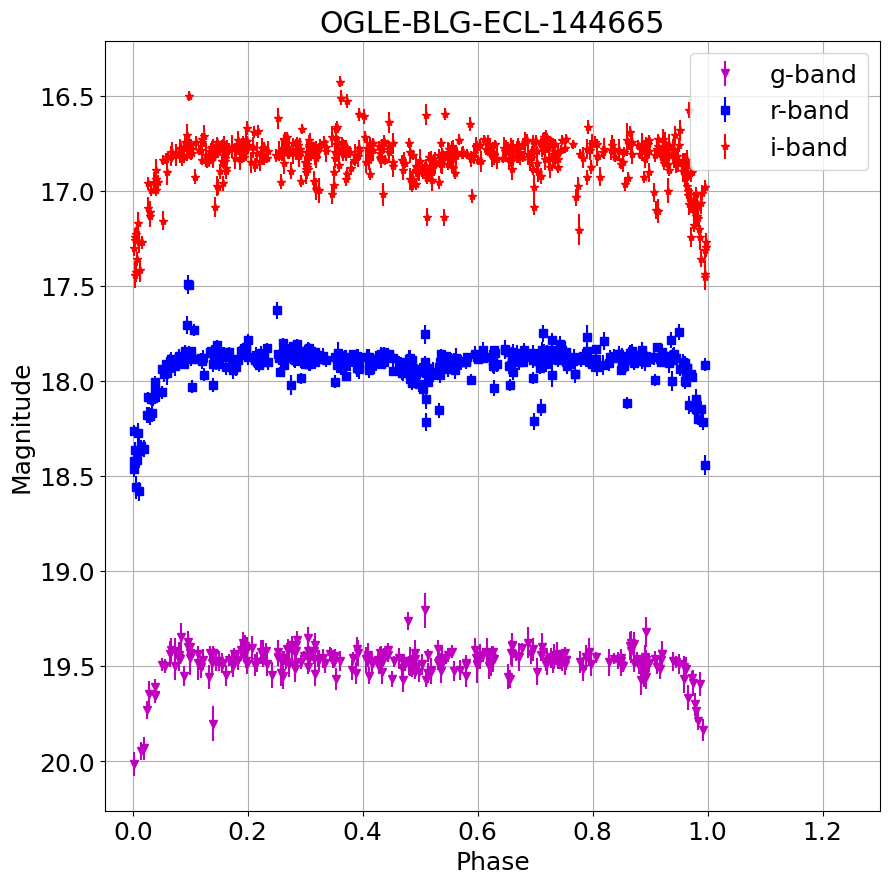} &
\includegraphics[width=0.45\textwidth]{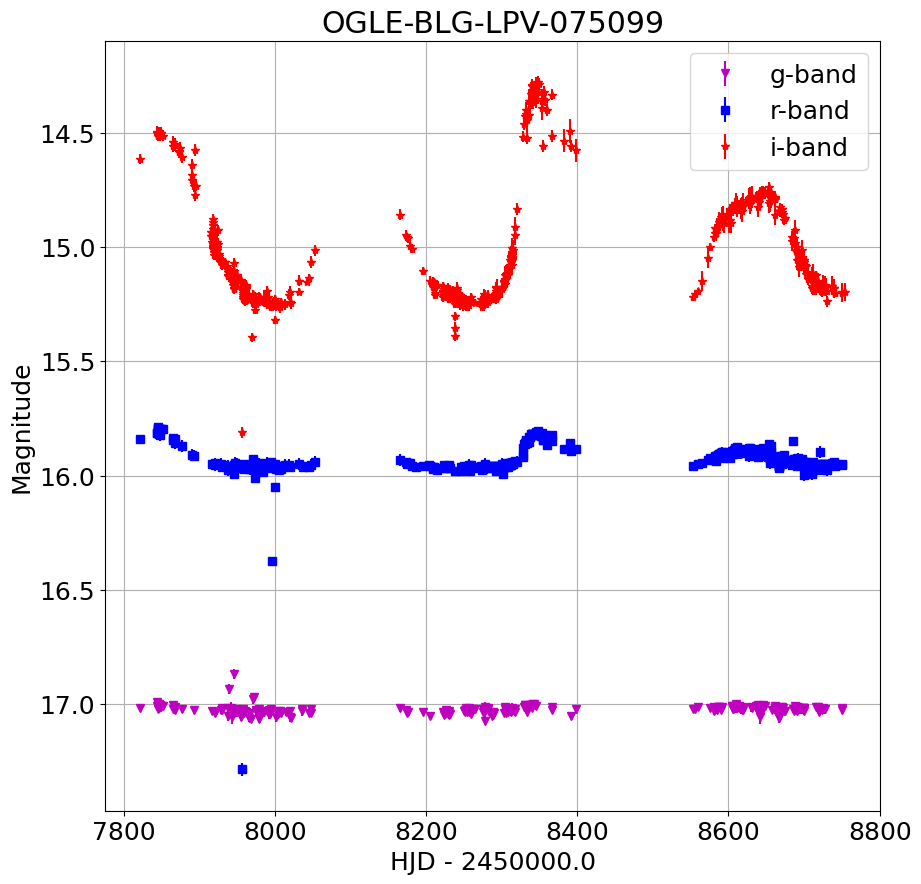}\\
\includegraphics[width=0.45\textwidth]{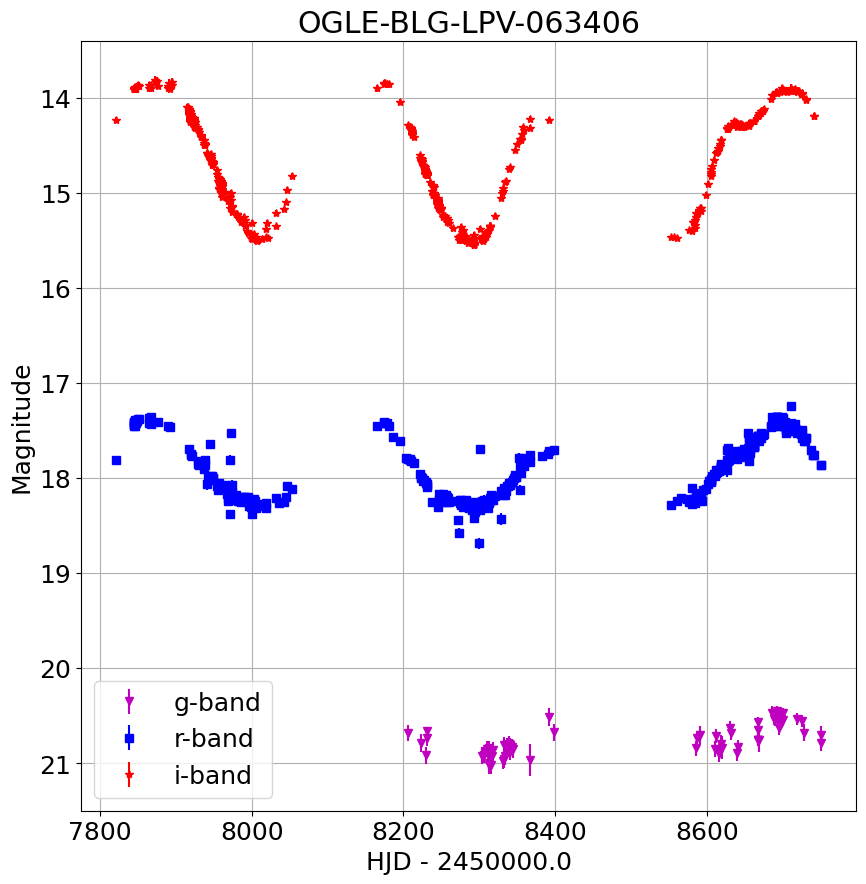} &
\includegraphics[width=0.45\textwidth]{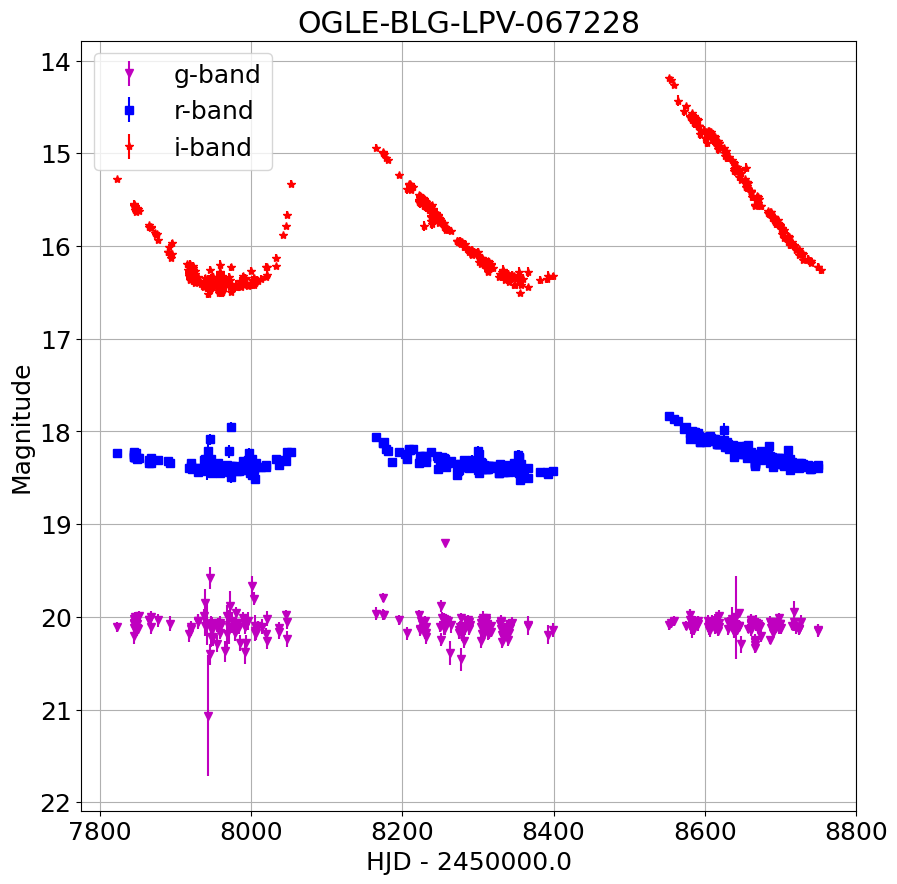} \\
\end{tabular}
\caption{A selection of light curves, filtered of inferior data, of variable stars from ROME-FIELD-01.\label{fig:variable_star_lcs}}
\end{centering}
\end{figure}

\begin{figure}
\begin{centering}
\includegraphics[width=0.8\textwidth]{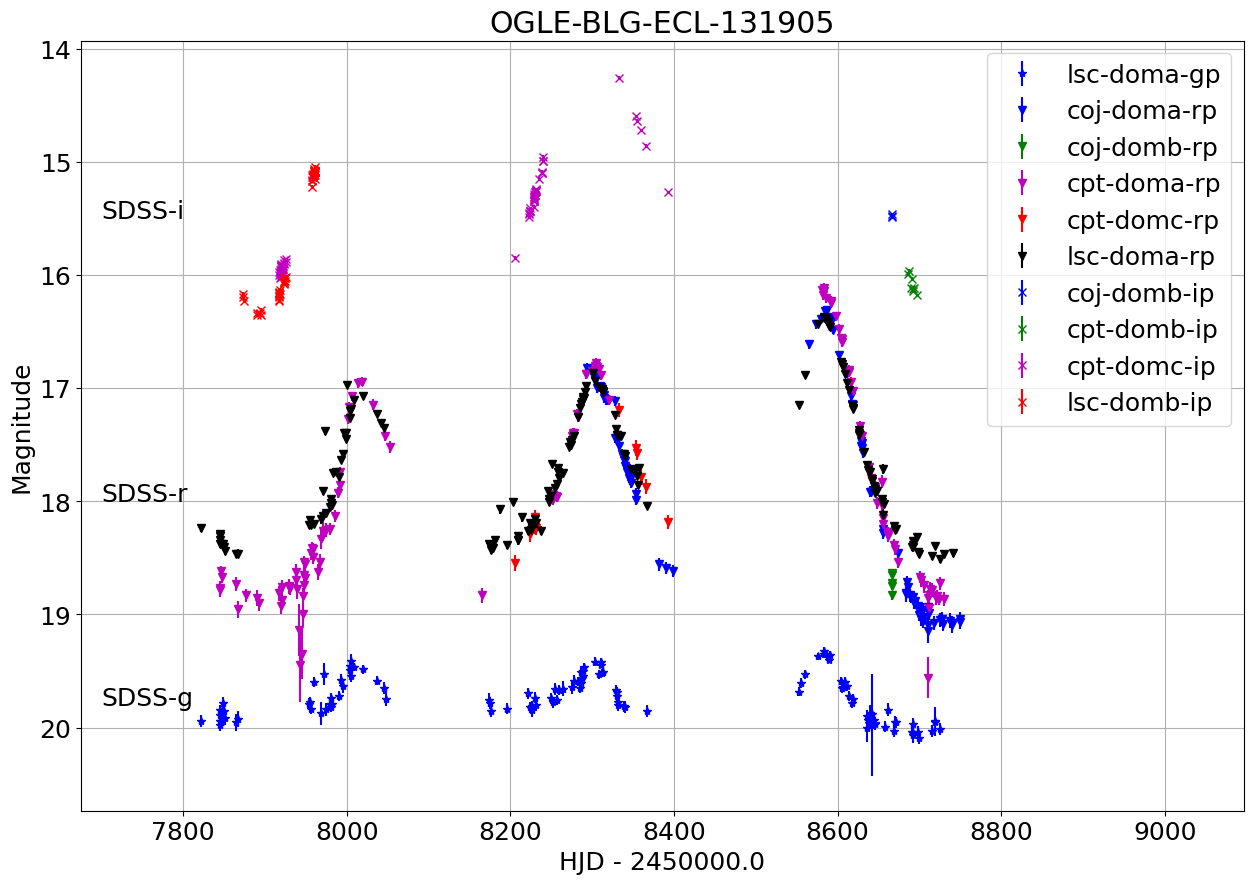} 
\caption{Light curve for OGLE-BLG-LPV-067512 (ROME-FIELD-01 star 208663), classified as a Mira, with the data from different telescopes and instruments plotted separately. 
  \label{fig:lc208663}}
\end{centering}
\end{figure}

In addition to microlensing events, the ROME survey provides long-baseline, 3-band lightcurves for all kinds of variable stars.  The ROME catalog was crossed matched against the OGLE Collection of Variable Stars \cite{Udalski2015b} and the VVV Variable Stars Catalog \cite{Molnar2022}, to facilitate the exploration of these data products for science with these objects.  Some examples of the lightcurve of both periodic and long-timescale variables are presented in Figure~\ref{fig:variable_star_lcs}, with the periodic lightcurves shown folded on the measured periods of the objects.  The contemporaneous observations in 3 passbands provides valuable time-variable color information but analyses should take care to account for blended flux.  The normalization process combines lightcurves  data from multiple sites as cleanly as possible, but when a variable star is blended with (usually non-variable) neighbors, the degree of blending can vary in the data from different sites at different times. It should be noted that the survey's overall photometric calibration is not designed to be absolute, since this is not required for our main science case.  Figure~\ref{fig:lc208663} shows an example of the data for a blended variable star, with the photometry from different sites and instruments distinguished, to illustrate the difference in measured amplitudes that results. 
More subtle effects of blending can also be seen in the lightcurves of the Long-Period Variables (LPVs) shown in Figure~\ref{fig:variable_star_lcs}.  The variation in this category of stars is normally due to stellar oscillations, which typically shows higher amplitudes in bluer passbands.  This is not reflected in the plotted lightcurves, although the correct identification of these stars in all passbands has been verified.  The dense crowding of this field means that all of these stars are blended to some degree.  From inspection of these stars in the reference images in each passband, the  elongation of the star in $i^{\prime}$-band is a telltail indication of blending that is less conspicuous or absent in $r^{\prime}$ and $g^{\prime}$.  This can mean that the flux measured in the $r^{\prime}$ and $g^{\prime}$ lightcurves can have a higher ratio of flux from the companion, or indeed come entirely from the compantion. It is therefore recommended that blending be taken into account during the modeling of variable star data, similar to the approach described above for microlensing events.  The ROME data products include information on the site/instrument origins of all datapoints to enable this analysis.  

%{\color{red}TBD: feedback welcome on what to include.  Suggestions so far:}
%\begin{itemize}
%    \item Example of the typical photometric precision achieved for the combined dataset photometry, e.g. RMS diagram for one field. Search for signs of systematic noise?
%    \item Curated catalog of known events with good coverage
%    \item CMD of field with cluster.  
%    \item Microlensing models for a few simple-to-model events, comparing the model parameters from ROME with those of, say, OGLE.
%    \item Table of known variables \& events.  This could be included in the IPAC source catalog.  This would increase the potential for others to use this dataset as a training dataset for ML. 
%    \item Suggest that applying LIA to the catalog is a separate paper. MPGH: Apply real/bogus, or indicate which stars are constant?
%\end{itemize}

% Necessary as we exceed the maximum number of available symbols for footnotes right about here
\setcounter{footnote}{0}

\section{IPAC Data Release Products}
\label{sec:data_products}
The ROME photometry catalog will be made publicly available through the NASA Exoplanet Archive\footnote{https://exoplanetarchive.ipac.caltech.edu/} hosted at the Infrared Processing and Analysis Center, IPAC.  These data products differ somewhat from the pyDANDIA field data products in order to make them compatible with IPAC standards.  This offers the advantage that the data can be visualized and explored through the {\em Firefly} framework\footnote{https://github.com/Caltech-IPAC/firefly}.  

The crossmatch tables for each field were combined to produce a single IPAC source catalog for the whole survey; the data included in this table for all stars is summarized in Tables~\ref{tab:ipac_source_table} -- \ref{tab:ipac_source_table_pt2}.  If stars corresponding to ROME sources were identified in cross-matches against other surveys, these are noted in the source catalog to facilitate the combination with other data products for microlensing and variable star science, and their use in training machine learning algorithms.  This includes microlensing event catalogs from OGLE, MOA and KMTNet, and variable source catalogs from OGLE, VVV and {\em Gaia} Alerts.  

The combined timeseries photometry is repackaged into one multi-extension FITS binary table file for each star.  The SDSS-$g^{\prime}$, $r^{\prime}$ \& $i^{\prime}$ lightcurves are included as sequential tables, and the data included for each lightcurve is summarized in Tables~\ref{tab:lc_header}--\ref{tab:lc_table}.  

The quality control index ({\texttt{qc\_flag}}) is assigned to each photometric measurement to indicate different issues with the data using a combination of bitmask values, summarized in Table~\ref{tab:qc_flags}.  Photometric residuals are calculated for each star's lightcurve by subtracting the star's mean magnitude weighted by the inverse variance of its photometric uncertainties.  If all stars in a given image exhibit unusually high residuals, all measurements from that image are flagged with \texttt{ qc\_flag}$+2$.   The photometric scale factor, produced by the Bramich algorithm is also used to evaluate measurement quality, normalized by the exposure time of each image, with \texttt{qc\_flag}$+4$.  Evaluating this metric for the whole ROME survey indicated that values $<$0.7 were a reliable indicator of poor photometric quality.  The transformation coefficients from the image alignment stage for all images are evaluated, and all photometry from images with outlier coefficients is assigned {\texttt qc\_flag}$+8$. Lastly, the median and standard deviation of the subtracted image stamps is calculated for all frames to flag the photometry from poor-quality subtractions with {\texttt qc\_flag}$+16$.  It should be noted that these {\texttt{qc\_flag}} does not flag images with high sky background or seeing, although these can result in poor quality photometry.  Instead, information is included in the lightcurves on lunar proximity and phase, airmass, sky background and PSF FWHM for all images, enabling the user to make their own selection cuts.  

\begin{table}
\begin{tabular}{|c|c|c|l|}
\hline
Column name           & Data type & Units & Description \\
\hline
 {\texttt{name}}       & string    & None  & Unique identifier assigned to the star, \\
                       & & & $<$field name$>$\_$<$field ID$>$\\
 {\texttt{field}} & string  & None & Name of the field the star is located in, \\
                        & & & format ROME-FIELD-XX, \\
                        & & & where XX = ${01…20}$\\
 {\texttt{field\_id}}  & integer   & None  & Star index number in the field\\
 {\texttt{ra}}         & double    & Decimal degrees & J2000.0 coordinates of the star\\
 {\texttt{dec}}        & double    & Decimal degrees & J2000.0 coordinates of the star\\
 {\texttt{quadrant}}   & integer   & None  & Quadrant number the star is assigned \\
                      &           &        & to in the field\\
 {\texttt{quadrant\_id}} & integer   & None  & Index of the star in the quadrant of\\
                      &            &        & the field\\
 {\texttt{gaia\_source\_id}} & integer & None & Source ID corresponding to the star, if \\
                     &              &       & any from {\em Gaia}-EDR3\\
 {\texttt{ogle\_event\_id}} & integer & None & OGLE event ID corresponding to the \\
                     &              &       & star, if any\\
 {\texttt{ogle\_variable\_id}} & integer & None & OGLE variable ID corresponding to the \\
                     &              &       & star, if any\\
 {\texttt{moa\_event\_id}} & integer & None & MOA event ID corresponding to the \\
                     &              &       & star, if any\\
 {\texttt{kmtnet\_event\_id}} & integer & None & KMTNet event ID corresponding to the \\
                     &              &       & star, if any\\
 {\texttt{spitzer\_event}} & boolean & None & Indicates if event observed by Spitzer\\
 {\texttt{vvv\_variable\_id}} & integer & None & VVV variable ID corresponding to the \\
                     &              &       & star, if any\\
 {\texttt{cal\_mag\_g}} & float    & Magnitude & Magnitude of the star in the survey \\
                    &               &       & reference dataset, calibrated to\\
                    &               &       & VPHAS+, SDSS-g\\
 {\texttt{cal\_mag\_error\_g}} & float    & Magnitude & Photometric uncertainty on the \\
                    &               &       & magnitude of the star in SDSS-g in the \\
                    &               &       & survey reference dataset\\
 {\texttt{norm\_mag\_g}} & float     & Magnitude & Magnitude of the star in SDSS-g, \\
                    &               &       & normalised to the primary reference \\
 {\texttt{norm\_mag\_error\_g}} & float & Magnitude & Photometric uncertainty of the star in \\
                    &               &       & SDSS-g, normalised to the primary \\
                    &               &       & reference dataset\\
 {\texttt{cal\_mag\_r}} & float    & Magnitude & Calibrated SDSS-r magnitude in the \\
                    &                &       & survey reference dataset\\
 \hline
\end{tabular}
\caption{Columns available in the ROME survey source catalog available through the NASA Exoplanet Archive \label{tab:ipac_source_table}}
\end{table}

\begin{table}[]
    \centering
    \begin{tabular}{|c|c|c|l|}
    \hline
    Column name           & Data type & Units & Description \\
    \hline
 {\texttt{cal\_mag\_error\_r}} & float    & Magnitude & Photometric uncertainty on the \\
                    &                &        & calibrated SDSS-r magnitude\\
 {\texttt{norm\_mag\_r}} & float     & Magnitude & Normalized magnitude in SDSS-r\\
 {\texttt{norm\_mag\_error\_r}} & float & Magnitude & Uncertainty on the normalized SDSS-r \\
                    &               &          & photometry\\
 {\texttt{cal\_mag\_i}} & float    & Magnitude & Calibrated SDSS-i magnitude in the \\
                    &                &       & survey reference dataset\\
 {\texttt{cal\_mag\_error\_i}} & float    & Magnitude & Photometric uncertainty on the \\
                    &                &        & calibrated SDSS-i magnitude\\
 {\texttt{norm\_mag\_i}} & float     & Magnitude & Normalized magnitude in SDSS-i\\
 {\texttt{norm\_mag\_error\_i}} & float & Magnitude & Uncertainty on the normalized SDSS-i \\
                    &               &          & photometry\\
 {\texttt{lc\_file\_path}} & string & None & Relative path to the lightcurve file\\
    \hline
    \end{tabular}
    \caption{Columns available in the ROME survey source catalog available through the NASA Exoplanet Archive (continued)}
    \label{tab:ipac_source_table_pt2}
\end{table}

\begin{table}
\begin{tabular}{|c|c|c|l|}
\hline
  Keyword & Data type & Units & Description  \\
\hline
 {\texttt{NAME}} & string & None & Unique identifier assigned to the star, \\
                & & & constructed from $<$field name$>$\_$<$field ID$>$\\
 {\texttt{FIELD}} & string & None & Name of the field the star is located in, of \\
                & & & the format ROME-FIELD-XX, \\
                & & & where XX = ${00…20}$\\
 {\texttt{FIELD\_ID}} & integer & None & Star index number in the field \\
 {\texttt{QUADRANT}} & integer & None & Field quadrant of that the star is assigned to\\
 {\texttt{QUAD\_ID}} & integer & None & Star index number in the field quadrant\\
 {\texttt{RA}}  & float & decimal degrees & Right Ascension of the star, J2000.0\\
 {\texttt{DEC}} & float & decimal degrees & Declination of the star, J2000.0\\
 {\texttt{GAIA\_ID}} & integer & None & {\em Gaia} source ID of the corresponding object \\
                & & & in the {\em Gaia} catalog \\
 {\texttt{GAIACAT}} & string & None & {\em Gaia} catalog version used to assign \\
                & & & a Gaia\_ID\\
 {\texttt{NDATA\_G}} & integer & None & Number of points in the SDSS-g lightcurve\\
 {\texttt{NDATA\_R}} & integer & None & Number of points in the SDSS-r lightcurve\\
 {\texttt{NDATA\_I}} & integer & None & Number of points in the SDSS-i lightcurve\\
\hline
\end{tabular}
\caption{Header keywords used for the Primary Header Data Unit for star lightcurves. \label{tab:lc_header}}
\end{table}

\begin{table}
\begin{tabular}{|c|c|c|l|}
    \hline
    Column name & Data Type & Units & Description \\
    \hline
    {\texttt{hjd}} & double & days & Heliocentric Julian Date of the photometric \\
                    &       &       & measurement\\
    {\texttt{inst\_mag}} & float & magnitudes & Instrumental magnitude photometric \\
                    &       &       & measurement\\
    {\texttt{inst\_mag\_error}} & float & magnitudes & Uncertainty on the instrumental magnitude \\
    {\texttt{calib\_mag}} & float & magnitudes & Calibrated reference image magnitude\\
    {\texttt{calib\_mag\_error}} & float & magnitudes & Uncertainty on the calibrated magnitude\\
    {\texttt{norm\_mag}} & float & magnitudes & Calibrated magnitude normalized to the \\
                         &       &              & primary reference dataset\\
    {\texttt{norm\_mag\_error}} & float & magnitudes & Uncertainty on the normalized magnitude\\
    {\texttt{qc\_flag}} & integer & None & Data quality control bitmask value\\
    {\texttt{dataset}} & string & None & Reference to the dataset from which the \\
                        &       &       & photometric datapoint is derived\\
    {\texttt{airmass}} & float  &  & Airmass \\
    {\texttt{moon\_frac}} & float &  & Moon phase during exposure, 0--1\\
    {\texttt{moon\_sep}} & float & Deg & Angular separation of the field center from \\
                        &        &      & the Moon during exposure\\
    {\texttt{sky\_bkgd}} & float & counts & Median sky background of exposure\\
    {\texttt{fwhm}} & float & pixels & Median Full-Width, Half-Maximum of the \\
                    &       &        & stellar Point Spread Function \\
        \hline
\end{tabular}
\caption{Columns in each FITS table extension for the timeseries photometry. \label{tab:lc_table}}
\end{table}

\begin{table}[]
    \centering
    \begin{tabular}{|c|l|}
    \hline
        {\texttt{qc\_flag}} & Description \\
    \hline
         0  & No known issue\\
         2  & Image photometry suffered above average residuals \\
         4  & Datapoint fails the photometric scale factor/exposure time criterion \\
         8  & Image had unreliable resampling coefficients \\
         16 & Poor quality image subtraction \\
    \hline
    \end{tabular}
    \caption{Bitmask values assigned to individual photometric points for different quality control issues during the reduction.}
    \label{tab:qc_flags}
\end{table}

\section{Concluding remarks and data usage policy}
\label{sec:concluding_remarks}
This data release is a public product of the ROME/REA Key Project survey conducted on the robotic telescope network of the Las Cumbres Observatory.

Individual images used in the data reduction are freely available for download on the LCO archive\footnote{https://archive.lco.global/}.

It is important to clarify that the light curves available through this data release are not optimized photometric reductions of individual objects. Researchers intending to utilize this data set for publications are kindly requested to acknowledge this source by citing the present paper along with the original work by Tsapras et al.~2019 \cite{Tsapras2019}. 

For research work that requires optimized photometry for specific targets within this catalog, we encourage reaching out to us directly.  Collaboration opportunities are sincerely welcomed, and we look forward to supporting further inquiries and investigations.

\section{Acknowledgements}
This work makes use of observations from the Las Cumbres Observatory global telescope network. RAS and EB gratefully acknowledge support from NASA grant 80NSSC19K0291. 
YT acknowledges the support of DFG priority program SPP 1992 “Exploring the Diversity of Extrasolar Planets” (TS 356/3-1). 
JW acknowledges the support of DFG priority program SPP 1992 "Exploring the Diversity of Extrasolar Planets" (WA 1047/11-1).  SM and WZ were partly supported by the National Science Foundation of China (Grant No. 12133005).  RFJ acknowledges support for this project provided by ANID's Millennium
Science Initiative through grant ICN12\textunderscore 009, awarded to the Millennium Institute of Astrophysics (MAS),
and by ANID's Basal project FB210003.
This paper made use of the tools and data provided by the NASA Exoplanet Database, together with many elements of the \texttt{astropy}
astronomical data analysis package, and the \texttt{Aladdin Sky Atlas} suite.  This work has made use of data from the European Space Agency (ESA) mission {\em Gaia} (\url{https://www.cosmos.esa.int/gaia}), procossed by the {\em Gaia} Data Processing and Analysis Consortium  (DPAC, \url{https://www.cosmos.esa.int/web/gaia/dpac/consortium}). Funding for the DPAC has been provided by national institutions, in particular the institutions participating in the {\em Gaia} Multilateral Agreement.  The ROME team would like to thank Michael Lund from the NASA Exoplanet Science Institute for his support in curating and archiving these data products for public release.  

\section*{References}

\bibliographystyle{unsrt}
\bibliography{references}
\end{document}